\documentclass[fleqn,usenatbib]{mnras}

\usepackage{newtxtext,newtxmath}

\usepackage[T1]{fontenc}

\DeclareRobustCommand{\VAN}[3]{#2}
\let\VANthebibliography\thebibliography
\def\thebibliography{\DeclareRobustCommand{\VAN}[3]{##3}\VANthebibliography}

\newcommand{\fracb}[2]{\left(\frac{#1}{#2}\right)}

\usepackage{graphicx}	
\usepackage{amsmath}	

\usepackage{adjustbox}
\usepackage{multirow}
\usepackage{tabularx}
\usepackage{booktabs}

\usepackage{lineno}

\usepackage{xcolor}  
\definecolor{blazeorange}{rgb}{1.0, 0.4, 0.0}
\definecolor{seagreen}{rgb}{0.18, 0.55, 0.34}
\definecolor{rufous}{rgb}{0.66, 0.11, 0.03}
\definecolor{royalfuchsia}{rgb}{0.79, 0.17, 0.57}
\definecolor{scarlet}{rgb}{1.0, 0.13, 0.0}
\definecolor{royalpurple}{rgb}{0.47, 0.32, 0.66}
\definecolor{darkblue}{rgb}{0, 0, 0.66}

\title[Afterglow Fits with Neural Emulation]{Very High-Energy GRB Afterglow Fits with Neural-Network Emulation and Bayesian inference }

\author[E. Aguilar-Ruiz et al.]{
Edilberto Aguilar-Ruiz,$^{1}$\thanks{E-mail: e.aguilar@irya.unam.mx}
Jessica Sald\'{i}var Talavera,$^{2}$
Ramandeep Gill,$^{1,3}$\thanks{E-mail: r.gill@irya.unam.mx}
Sundar Srinivasan$^{1}$
\\
$^{1}$Instituto de Radioastronom\'ia y Astrof\'isica, Universidad Nacional Aut\'onoma de M\'exico, Antigua Carretera a P\'atzcuaro $\#$ 8701,  Ex-Hda. San Jos\'e de la \\ Huerta, Morelia, Michoac\'an, C.P. 58089, M\'exico\\
$^{2}$ Escuela Nacional de Estudios Superiores Unidad Morelia, Universidad Nacional Aut\'onoma de M\'exico, Antigua Carretera a Pátzcuaro No. 8701, Col. Ex \\ Hacienda de San José de la Huerta, C.P. 58190 Morelia, Michoacán, México \\
$^{3}$ Astrophysics Research Center of the Open university (ARCO), The Open University of Israel, P.O Box 808, Ra'anana 43537, Israel
}

\date{Accepted XXX. Received YYY; in original form ZZZ}

\pubyear{\the\year{}}

\begin{document}


\label{firstpage}
\pagerange{\pageref{firstpage}--\pageref{lastpage}}
\maketitle

\begin{abstract}
The very high-energy ($E > 100$\,GeV) afterglow emission in gamma-ray bursts (GRBs) can be explained by synchrotron self-Compton (SSC). Modeling this emission accurately requires fully numerical codes that solve the time-dependent coupled particle--photon kinetic equations while accounting for Klein--Nishina suppression. However, their substantial computational cost makes them impractical for traditional parameter inference techniques such as Markov Chain Monte Carlo (MCMC) sampling. In this work, we present a highly efficient surrogate model based on artificial neural networks (ANNs) that emulates a one-zone kinetic SSC radiation code. To train our ANN, we use a multi-dimensional model parameter grid that constitutes a sample of 100,800 spectra produced by an infinitely thin spherical blast wave propagating inside a constant density interstellar medium (ISM). The ANN reproduces the numerical spectra well, with the median relative error generally below 2\,per cent across the full frequency domain. Using this surrogate framework, we perform Bayesian parameter posterior estimation via MCMC and fit the broadband GRB afterglow observations of GRB\,190114C from X-rays to TeV $\gamma$-rays at different epochs. We find an energetic jet with isotropic-equivalent kinetic energy of $E_{\rm k,iso}\simeq7.2\times10^{54}$\,erg moving with a coasting bulk Lorentz factor of $\Gamma\gtrsim500$ into an ISM with density $n\simeq0.1\,{\rm cm^{-3}}$. The deviation of our model light curve from observations on long timescales suggests that an ISM environment is inconsistent, and instead favours a radially stratified external medium.
\end{abstract}




\section{Introduction}
The long-lasting afterglow emission of gamma-ray bursts (GRBs) is detected across multiple frequency bands, from radio to gamma rays. In the fireball model \citep{Goodman-86, Paczynski-86, Shemi-Piran-90} this emission arises when the ultra-relativistic outflow, with bulk Lorentz factor (LF) $\Gamma\sim10^2-10^3$, launched by a central compact engine interacts with the circumburst environment \citep{Rees-Meszaros-92,Meszaros-Rees-93,Meszaros-Rees-97}, i.e., an interestellar (ISM) or a stellar-wind medium. This interaction produces two external shocks, where a forward shock decelerates by sweeping up the circumburst material while a reverse shock decelerates the ejecta \citep{Sari-Piran-95}. While synchrotron radiation from non-thermal electrons accelerated at the forward shock can well explain the broadband emission up to $\sim$MeV-GeV $\gamma$-rays \citep{Sari+98, Granot-Sari-02}, theoretical arguments suggest that it cannot account for emission above the maximum synchrotron energy of $E_{\rm syn,\max} \simeq 7(1+z)^{-1}\kappa\Gamma_2$\,GeV \citep{Guilbert+83,deJager-Harding-92}, where $\kappa$ quantifies the acceleration efficiency and $\Gamma_2=\Gamma/100$.

Afterglow observations at very-high energies (VHE; $E>100$\,GeV) in, e.g., GRB\,190114C \citep{MAGIC_GRB190114C}, GRBs 160821B \citep{Acciari+21}, 180720B \citep{Abdalla+19}, 190829A \citep{HESS_COLLABORATION+21}, 201216C \citep{Fukami+22}, and most notably from the brightest GRB observed to date, GRB\,221009A \citep{LHAASO+23}, confirm that GRB afterglow emission can extend well beyond the theoretical synchrotron limit. These observations demonstrate that mechanisms, such as synchrotron self-Compton (SSC) or external Compton (EC), are required to explain the highest-energy photons. Furthermore, detection of VHE emission from GRBs provides crucial insight into particle acceleration processes and the microphysics of relativistic shocks, offering a unique probe of extreme astrophysical conditions. 

In the SSC framework, synchrotron photons produced by relativistic electrons are upscattered to higher energies via inverse Compton scattering by the same electron population \citep[e.g.][]{Panaitescu-Meszaros-98, Dermer+00a, Dermer+00b, Panaitescu-Kumar-00}. This process naturally extends the afterglow emission into the VHE regime and can produce a spectral transition at GeV energies, where the declining synchrotron component overlaps with the rising SSC contribution. Such features have been observed in a couple of TeV detected GRBs \citep[i.e.,][]{MAGIC_GRB190114C, Banerjee+25}.

Modeling of the VHE emission of GRB afterglows has been widely performed using analytic SSC approaches \citep[e.g.][]{Sari-Esin-01, Nakar+09, Jacovich1+21, McCarthy-Laskar-24, Yamasaki-Piran-22}. These model offer high computational efficiency and their relatively easy to implement framework makes them well suited for parameter inference using techniques such as Markov Chain Monte Carlo (MCMC) sampling. 
However, these models are often limited in accuracy due to reliance on simplifying assumptions, namely restriction to emission only along the line of sight (LOS), adoption of sharply broken power-law particle spectra, usage of a step-function approximation for the Compton cross section, and a simplified radiative transfer treatment that assumes instantaneous photon escape and neglects adiabatic cooling.

More sophisticated numerical models relax many of these assumptions by incorporating a more realistic treatment of radiative losses and time-dependent photon/particle evolution by solving the kinetic photon-particle coupled equations \citep[e.g.][]{Petropoulou-Mastichiadis-09, Derishev-Piran-21, Nedora+24, Hope+25}, but at the cost of high computational demand making them less practical for model fitting and parameter inference. Semi-analytic approaches, such as the method proposed by \citet{Aguilar-Ruiz+26}, bridge the gap between analytic and fully numerical models, reproducing the broadband afterglow spectrum and its temporal evolution in a manner comparable to kinetic approaches. Nevertheless, despite these improvements, its computational cost remains higher than that of purely analytic models and may become impractical when more complex physical ingredients, such as jet structure or reverse shocks, are included.

Despite the need for physically accurate SSC modeling, the inclusion of more detailed and realistic assumptions makes efficient parameter inference computationally challenging. This limitation is particularly relevant in the context of contemporary multiband and VHE GRB afterglow observations, which require extensive exploration of parameter space and robust statistical constraints. In this context, Artificial Neural Network (ANN) emulators \citep[e.g.][]{Gupta-Singh-2002, Smith-Geach-2023} offer an efficient alternative to drastically reduce computational cost while retaining high accuracy similar to full numerical treatment. An ANN learns from precomputed numerical simulations to predict the outcome for sets of parameters not included in the training sample. In this paper, the ANN acts as a forward-model emulator by mapping physical model parameter to predict spectra; it does not directly approximate the posterior distribution. Once trained, the ANN can emulate complex numerical models with reasonably high accuracy, but orders of magnitude faster than the original simulations. ANNs have been applied in many astrophysical contexts, e.g., in cosmological applications \citep[e.g., ][]{Jamieson+23, Cabayol-Garcia+23, Jin+25, Palud+23}, and the modeling of broadband blazar emission \citep[e.g.,][]{Begue+24, Sahakyan+24,Sahakyan+25, Tzavella+24}. Nevertheless, in the context of GRB afterglows, ANNs have only been considered for synchrotron emission \cite[e.g.,][]{Boersma+23}, based on analytical emission model provided by the \texttt{BOXFIT} code \citep{vanEerten+12} as training dataset. 

In this work, we present an ANN-based surrogate model that emulates a numerical SSC afterglow model. 
To construct the training grid, we adopt the kinetic code developed in \citet{Gill-Thompson-14} to model the prompt GRB emission \citep{Vianello+18,Gill-Granot-18a,Gill+20b} and use it to simulate the SSC afterglow emission in the (comoving) frame of the shocked fluid.
To calculate the broadband spectrum in the observer frame we assume a spherical flow in the thin-shell approximation propagating into a constant density ISM, and calculate the observed flux by integrating over the equal-arrival-time-surface (EATS).  We adopt the ISM environment as a first application of the proposed ANN-based emulation framework. Although the methodology is not restricted to this particular environment, extending the training grid to include alternative external density profiles would substantially increase the computational cost of generating the corresponding model grid.
We find that the resulting ANN surrogate drastically reduces the computational cost of forward-modeling evaluations, enabling efficient TeV GRB afterglow modeling and likelihood-based Bayesian parameter inference via MCMC sampling, while retaining accuracy similar to full numerical models.

The structure of this paper is as follows. In Section \ref{sec:the_model}, we describe the physical model adopted for GRB afterglows. Section \ref{sec:radiation_code} describes the kinetic approach, its numerical implementation, and the underlying
physical assumptions. Section \ref{sec:NN&MCMC} describes the construction and validation of the ANN surrogate model and the MCMC-based Bayesian inference framework. Section \ref{sec:model_fits} presents the 
application of this framework to broadband GRB afterglow observations and the resulting constraints on the parameters.
Finally, in section \ref{sec:Discussion} we discuss the results.

\section{Afterglow Model}\label{sec:the_model}

\subsection{Spherical Thin Shell Dynamics}
We consider the dynamical evolution of an ultrarelativistic thin spherical shell ejected by the central engine. This mass shell has baryonic mass $M_{\rm ej}$ and an isotropic-equivalent kinetic energy $E_{k,\rm iso} = (\Gamma_0-1)M_{\rm ej}c^2$, with $c$ being the speed of light, where it initially moves with a coasting bulk Lorentz factor of $\Gamma_0\gg 1$. The external medium into which it propagates may have a radial density profile with $\rho(R)=m_p n(R) = AR^{-k}$, where $n$ is the number density, $m_p$ is the proton mass, $A = m_pn_0R_0^k$, and $n_0$ is the number density normalization at a radius $R_0$ from the central engine. For a constant density interstellar medium (ISM), which is valid for short GRBs, $k=0$, and in the case of long GRBs both an ISM or a wind profile with $k=2$ are typically assumed. The shell begins to sweep up the external medium in its path, causing it to slow down, and after the swept up mass, $M_{\rm sw}(R) = 4\pi\rho(R)R^3/(3-k)$, reaches a critical value of $M_{\rm ej}/\Gamma_0$, the dynamical evolution of the shell becomes self-similar \citep{Blandford-McKee-76}. This transition occurs at the deceleration radius, 
$R_{\rm dec} = \left[(3-k)E_{k,\rm iso}/4\pi Ac^2\Gamma_0^2\right]^{1/(3-k)}$, at which point most of the kinetic energy of the shell is transferred to the internal and kinetic energy of the swept up external medium. 

The interaction of the shell with the external medium gives rise to a two shock structure, 
where the \textit{forward} shock, with $\Gamma_{\rm sh}\approx\sqrt{2}\Gamma$ when $\Gamma\gg1$, 
shock heats the swept up external medium, while the \textit{reverse} shock propagates into 
the shell and shock heats it and slows it down. The shocked material behind the two shocks 
is separated by a contact discontinuity where an approximate pressure equilibrium is maintained 
between two shocked materials. Here we make a simplifying assumption and ignore the radial structure 
of the two shocks and instead assume that the entire shocked shell moves at a common bulk-$\Gamma$, which at $R>R_{\rm dec}$ follows a power law profile in radius \citep[$\Gamma\propto R^{-(3-k)/2}$;][]{Panaitescu-Kumar-00,Gill-Granot-18b}, such that
\begin{equation}
\label{eq:Gamma}
    \Gamma(\xi) = \frac{\Gamma_0+1}{2}\xi^{k-3}\left[\sqrt{1+\frac{4\Gamma_0}{\Gamma_0+1}\xi^{3-k}+\left(\frac{2\xi^{3-k}}{\Gamma_0+1}\right)^2}-1\right]\,,
\end{equation}
where $\xi\equiv R/R_{\rm dec}$. The above solution is valid from the ultrarelativistic regime to the trans-relativistic regime when $\Gamma\gtrsim1$. We make another approximation where we only consider the bulk-$\Gamma$ of the shocked material behind the forward shock, as given by Eq.\,(\ref{eq:Gamma}), to calculate both the dynamics of and radiation from the shocked shell.

\subsection{Synchrotron Emission from the Forward Shock}
The external medium upstream of the forward shock is assumed to be cold with enthalpy density given simply by its rest mass energy density, $w=m_pn(R)c^2$. As this material enters the forward shock, it is shock-heated and acquires an internal energy density\footnote{All primed quantities henceforth are expressed in the comoving (fluid) frame.}, $e' = (\Gamma-1)n'm_pc^2$, where the material just behind the shock has bulk Lorentz factor $\Gamma$. The shock also compresses the swept up material to a number density $n'=(\hat\gamma\Gamma+1)/(\hat\gamma-1)n = 4\Gamma n$, where the last equality is true when the adiabatic index of the shocked material is given by $\hat\gamma=(4\Gamma+1)/3\Gamma$, which switches from the relativsitic ($\hat\gamma=4/3$) to the non-relativistic ($\hat\gamma=5/3$) regime as the shock slows down. 

According to the standard model of afterglow emission, a fraction $\epsilon_e$ of the 
energy density of the shocked material goes into accelerating a fraction $\xi_e$ of the 
total number of electrons behind the shock into a power-law energy distribution, with 
$dn'/d\gamma\propto\gamma^{-p}$ with a power-law index $p$, and for $\gamma_m \leq \gamma \leq \gamma_M$ where $\gamma$ 
is the electron Lorentz factor. The minimum Lorentz factor of the power-law distribution 
is given by $\gamma_m = (\epsilon_e/\xi_e)[(p-2)/(p-1)](m_p/m_e)(\Gamma-1)$ for $2\lesssim p \lesssim3$, 
and the maximum Lorentz factor $\gamma_M$ is set by the balance between particle acceleration and 
cooling at the shock. The remaining $(1-\xi_e)$ electrons form a thermal distribution with a mean Lorentz factor 
$\langle\gamma_{\rm th}\rangle\lesssim\gamma_m$. 
A further fraction $\epsilon_B$ of the internal energy goes into self-generating, and/or amplifying 
any pre-existing magnetic fields in the upstream material into, small-scale random magnetic 
fields behind the shock with strength $B' = \left[32\pi\Gamma(\Gamma-1)\epsilon_Bnm_pc^2\right]^{1/2}$. 

The power-law electrons will gyrate around this small-scale B-field and cool by emitting 
synchrotron radiation at a characteristic frequency $\nu_{\rm obs} \approx (1+z)^{-1}\Gamma\nu_{\rm syn}'$, 
where $\nu_{\rm obs}$ is the frequency in the observer-frame, $z$ is the source redshift, and 
$\nu_{\rm syn}'(\gamma) \approx \gamma^2q_eB'/2\pi m_ec$ with $q_e$ being the elementary charge. 
The spectrum of this emission is non-thermal and comprises multiple 
power-law segments joined smoothly at characteristic break frequencies \citep{Sari+98,Granot-Sari-02}, 
namely the self-absorption frequency ($\nu_{\rm sa}$), the cooling break ($\nu_c$), and the characteristic synchrotron frequency of minimal energy electrons ($\nu_m$). The relative ordering of $\nu_m$ and $\nu_c$ determines whether the emitting electrons are in the \textit{fast cooling} regime, in which all electrons are cooling on a timescale shorter than the dynamical time, or in the \textit{slow cooling} regime, for which electrons with $\gamma<\gamma_c$ remain uncooled and ones with $\gamma>\gamma_c$ have lost most of their kinetic energy to synchrotron radiation. The cooling Lorentz factor, $\gamma_c$, is typically obtained by equating the synchrotron cooling time, $t_{\rm syn}' = 6\pi m_ec/\sigma_TB'^2\gamma$, to the dynamical time, $t'_{\rm dyn} = R/\Gamma\beta c$, to obtain $\gamma_c = (6\pi m_ec^2/\sigma_T)(\Gamma\beta/B'^2R)$. This is a global approximation that only accounts for the properties of the shocked material just behind the shock and ignores the radial dependence of the shocked fluid properties downstream of the shock. Local cooling treatments \citep{Granot+99,Granot-Sari-02} that explicitly account for this radial dependence find the cooling break frequency to be larger by a factor of few to several tens \citep{van-Eerten+2010,Kundu-van-Eerten-2026}.

\subsection{Synchrotron Self-Compton Radiation}
In addition to the standard synchrotron emission, electrons accelerated by the forward shock unavoidably up-scatter synchrotron photons via the inverse Compton scattering under the known synchrotron self-Compton (SSC) mechanism. This process naturally produces a high-energy component that extends to $\gamma$-rays with energy above $\sim \rm GeV$. The SSC emission is relevant for interpreting VHE observations of GRB afterglows since such emission cannot be explained by the synchrotron mechanism alone. 

The peak flux of the SSC component compared to that of synchrotron 
is obtained with the Compton-Y parameter, such that $\nu_{\rm ssc} F_{\nu, \rm ssc} \sim Y_{\rm ssc}  \, \nu_{\rm syn} F_{\nu, \rm syn}$, which in the Thomson regime mainly depends on the ratio of microphysical parameters, $Y_{\rm ssc } (r) \sim \sqrt{ \eta_{\rm rad}(r) \epsilon_e/\epsilon_B}$ \citep[e.g.][]{Moderski+00,Sari-Esin-01}. Its evolution is determined by the conditions of the external medium and the forward shock dynamics through the SSC radiative efficiency given by $\eta_{\rm rad} = \left( \gamma_m/\gamma_c\right)^{p-2}$. The Compton-Y parameter affects the cooled particle distribution as $n' \propto \frac{1}{1+Y_{\rm ssc}}\gamma^{-p-1}$ \citep[e.g.][]{Nakar+09}, consequently reducing the overall synchrotron fluxes and shape.

At very high energies, when the photon energy in the electron rest frame exceeds the electron rest-mass energy, i.e., $h\nu' \gamma \gtrsim m_e c^2$, the SSC emission is strongly suppressed due to a reduction in the scattering cross section caused by Klein--Nishina (KN) effects. Accounting for KN effects is crucial when modeling GRB afterglows, as they also modify the shape of the seed synchrotron spectrum \citep{Nakar+09,Jacovich1+21,McCarthy-Laskar-24}. Recently, \citet{Aguilar-Ruiz+26} demonstrated that a 
more accurate calculation of the Compton-$Y$ parameter, and hence a more realistic description of the afterglow phase within the SSC framework, requires the inclusion of adiabatic expansion and photon escape effects. These processes are important because they dilute the number density of synchrotron photons within the emitting shell, consequently reducing the efficiency of Compton scattering.

\section{One-Zone Radiation Code}\label{sec:radiation_code}
To calculate the local emission more accurately than the canonical assumption of a broken power-law comoving spectrum arising from a purely power-law particle distribution, we solve the coupled 
kinetic equations for both the electrons and photons in a one-zone approximation. This approximation is particularly suited for a thin-shell that ignores any radial profile of the shocked material and assumes that all of the emission is produced in an infinitely thin layer immediately behind the shock. Furthermore, since the shell is assumed to be uniform, a single zone or a spherical \textit{blob} is used to approximate the emission from the different angular regions of the flow. Most importantly, the one-zone approximation implies that the size of the simulated emission region is as large as the causal size of the flow, i.e. $R/\Gamma$ in the comoving frame. The implication of this assumption is that all particles injected into the zone interact with all photons produced inside this zone. Moreover, for simplicity, both species are treated as isotropic and so is their mutual interaction.

The radiation code solves the following two coupled equations in the comoving frame \citep{Peer-Waxman-05,Belmont+08,Vurm-Poutanen-09,Gill-Thompson-14,Gill+20b} for electrons and photons,
\begin{eqnarray}
    \frac{\partial^2 n_e'(p)}{\partial p\,\partial t'} &&= \dot n_{e,\rm syn}' + \dot n_{e,\rm IC}' + 
    \dot n_{e,\rm ad}' + \dot n_{e,\rm coul} + \dot n_{e,\rm inj} \\
    \frac{\partial^2 n_\gamma'(x')}{\partial x'\, \partial t'} &&= \dot n_{\gamma,\rm syn}' + \dot n_{\gamma,\rm IC}' + 
    \dot n_{\gamma,\rm ad}' + \dot n_{\gamma,\rm esc}\,, \\
\end{eqnarray}
over a logarithmic grid of dimensionless particle momenta $p\equiv\gamma\beta_e$, with $\beta_e=\sqrt{1-\gamma^{-2}}$, and dimensionless photon energies $x'\equiv h\nu'/m_ec^2$, where $h$ is Planck's constant. The right hand side represents a sum of reaction rates of all the different radiative processes due to which the particle distribution and photon spectrum evolves both in energy/momentum and time, namely synchrotron emission and self absorption (syn), inverse Compton (IC), adiabatic cooling and density dilution (ad), Coulomb interactions (coul), and escape from the emission region (esc). Coulomb interactions are only valid for electrons and only radiation is allowed to escape whereas the electrons are assumed to remain inside the emission region. 

\subsection{Integro-Differential Split Formalism}
To preserve accuracy and resolve energy and momentum transfers at the sub-grid level, we adopt a \textit{split} formalism that solves the evolution equations of particles and photons in two different regimes \citep[see, e.g.,][for further discussion]{Belmont+08,Belmont-09}. When the outgoing photon/particle receives a large energy/momentum change with respect to the incoming particle/photon, the interactions can be resolved over the energy/momentum grid and are described by exact collision integrals. For example, photons undergoing inverse Compton scattering in which process the mean energy of the scattered photon is given by $\nu'\sim\gamma^2\nu_0'$, with $\nu_0'$ being the energy of the incoming photon, are evolved using the collision integrals. On the other hand, when such energy/momentum transfers are smaller than the grid scale, a more accurate treatment involves the use of Fokker-Planck (FP) equations \citep{Nayakshin-Melia-98}

\vspace{-0.4cm}
\begin{equation}
    \dot n_{e,\rm FP}'(p) =\partial_p\left[\frac{\gamma}{p}A_e(p)n_e'(p)\right] 
    + \frac{1}{2}\partial_p\left[\frac{\gamma}{p}\partial_p\left\{\frac{\gamma}{p}D_e(p)n_e'(p)\right\}\right] \, ,
\end{equation}
\vspace{-0.9cm}
\begin{equation}
    \dot n_{\gamma,\rm FP}'(x') =\partial_{x'}\left[A_\gamma(x') n_\gamma'(x')\right] + \frac{1}{2}\partial_{x'}^2\left[D_\gamma(x') n_\gamma'(x')\right]\,.
\end{equation}
The advection coefficients $A_e(p)$ and $A_\gamma(x')$ represent the mean rate of change of particle/photon energy due to, e.g., secular heating/cooling processes and expansion, and the corresponding diffusion coefficients are given by $D_e(p)$ and $D_\gamma(x')$. This formalism is used to treat Coulomb interactions between electrons (and also electron-positron pairs if they were to become important at earlier stages of the flow), Compton scattering, synchrotron emission and self-absorption, and adiabatic cooling due to expansion. The full evolution of the photon spectrum and particle distribution is then obtained from a combination of the FP equations and collision integrals \citep[see][for more details]{Belmont+08},
\begin{equation}
    \partial_{t'}\{n_e'(p),n_\gamma'(x')\} = \partial_{t'}\{n_e'(p),n_\gamma'(x')\}^{\rm FP} + \partial_{t'}\{n_e'(p),n_\gamma'(x')\}^{\rm col}\,.
\end{equation}

\subsection{Electron Injection}
To calculate the emission using the numerical formalism discussed in \S\ref{sec:radiation_code}, it is essential to know the rate at which electrons are injected into the shocked region. The rate of injection per unit comoving volume can be expressed as 
\begin{equation}
    \dot n' = \frac{1}{V'}\frac{dN'}{dt'}\,,
\end{equation}
where $V' = 4\pi R^2\Delta'$ is the volume of the shocked region and $\Delta'$ is its radial width. Particle number conservation dictates that the number of particles swept up by the forward shock, $N = M_{\rm sw}(R)/(m_p+m_e)\approx M_{\rm sw}(R)/m_p$, must be equal to that behind the shock, $N' = n'V'$, which yields $\Delta' = R/[4(3-k)\Gamma]$. Likewise, for the injection rate $dN' = dN = n(R)dV = 4\pi n(R)R^2dR$ and $dR = \Gamma\beta cdt'$, which then yields
\begin{equation}
    \dot n' = \frac{4(3-k)n\Gamma^2\beta c}{R}\,.
\end{equation}
The time dependent power-law energy distribution of the non-thermal electrons injected at the forward shock is given by
\begin{equation}
    \dot n_{e,\rm inj}'(\gamma) = \frac{dn_{e,\rm inj}'}{dt'd\gamma} = A_{e,\rm inj}\gamma^{-p}\,,\quad\quad \gamma_m\leq\gamma\leq\gamma_M\,.
\end{equation}
If a fraction $\xi_e$ of the injected electrons are accelerated into a power-law energy distribution, then 
\begin{equation}
    \xi_e\dot n' = \int_{\gamma_m}^{\gamma_M}\dot n_{e,\rm inj}'(\gamma)d\gamma
\end{equation}
which yields the normalization
\begin{equation}
    A_{e,\rm inj} = \frac{(1-p)}{\gamma_M^{1-p}-\gamma_m^{1-p}}\xi_e\dot n'\,.
\end{equation}
The injected distribution over particle momenta $p$ instead of $\gamma = \sqrt{1+p^2}$ can now be obtained by simply multiplying by the Jacobian of transformation, 
\begin{equation}
    \dot n_{e,\rm inj}'(p) = \frac{d\gamma}{dp}\dot n_{e,\rm inj}'(\gamma) = \frac{p}{\gamma}\dot n_{e,\rm inj}'(\gamma)\,.
\end{equation}
At any given radius $R$, the particle number density is $n'=4\Gamma n$ and the Thomson optical depth associated with this number density in the causal region is 
\begin{equation}
    \tau_T = n'\sigma_T\frac{R}{\Gamma} = 4\sigma_TnR \simeq 2.7\times 10^{-7}n_0R_{18}\,,
\end{equation}
where $\sigma_T$ is the Thomson cross-section. The above estimate suggests that the emission region remains optically thin during the afterglow.

\subsection{Adiabatic Cooling \& Density Dilution due to Expansion}
In general, particles confined to an expanding volume lose internal energy by doing $PdV$ work. If the number of particles does not increase at the same rate as the volume then their density must also decline.  The equation governing adiabatic losses is given as \citep[e.g.][]{Longair+11}
\begin{equation}
    \frac{\partial n'(E',t')}{\partial t'} = \frac{\partial}{\partial E'}\left[-\frac{dE'}{dt'}n'(E',t')\right]-\frac{d\ln V'}{dt'}n'(E',t')\,,
\end{equation}
where $n'$ is the number density of particles with energy $E'$ that are enclosed inside a volume $V'$. The first term on the right hand side gives the adiabatic cooling and the second gives the density dilution. To understand how the particle's energy is altered adiabatically as the confining volume is changed, we start with the first law of thermodynamics that states $dU' = -PdV'$, where $U' = N'\langle E'\rangle$ is the total internal energy of $N'$ particles with mean energy $\langle E'\rangle$ and $P = (\hat\gamma-1)e'$ is the gas pressure that is related to the internal 
energy density $e'=U'/V'$ through the adiabatic constant $\hat\gamma$. Substituting the expression for pressure and energy density, and if the number of particles is conserved, then the thermodynamic equation can be expressed as $d\ln E' = -(\hat\gamma-1)d\ln V'$. Differentiating both sides with respect to the comoving time gives
\begin{equation}
    \frac{d\ln E'}{dt'} = -(\hat\gamma-1)\frac{d\ln V'}{dt'} = -\frac{(\hat\gamma-1)}{t_{\rm ad}'}\,,
\end{equation}
where $t_{\rm ad}'$ is the adiabatic cooling time. As the flow expands, its comoving volume changes according to
\begin{equation}
    \frac{d\ln V'}{dR} = \frac{3}{R}\left(1-\frac{1}{3}\frac{d\ln\Gamma}{d\ln R}\right)\,.
\end{equation}
Substituting $dR = \Gamma\beta cdt'$ in the above equation then yields the rate of change of particle energy
\begin{equation}
    \frac{d\ln E'}{dt'} = -(\hat\gamma-1)\frac{3\Gamma\beta c}{R}\left(1-\frac{1}{3}\frac{d\ln\Gamma}{d\ln R}\right)\,.
\end{equation}
In the coasting phase, $d\ln\Gamma/d\ln R = 0$, and in the self-similar phase for $R_{\rm dec}<R<R_{\rm NR}$, where $R_{\rm NR}$ is the non-relativistic transition, the asymptotic $d\ln\Gamma/d\ln R = -(3-k)/2$.

\begin{figure*}
    \centering
    \includegraphics[width=0.33\textwidth]{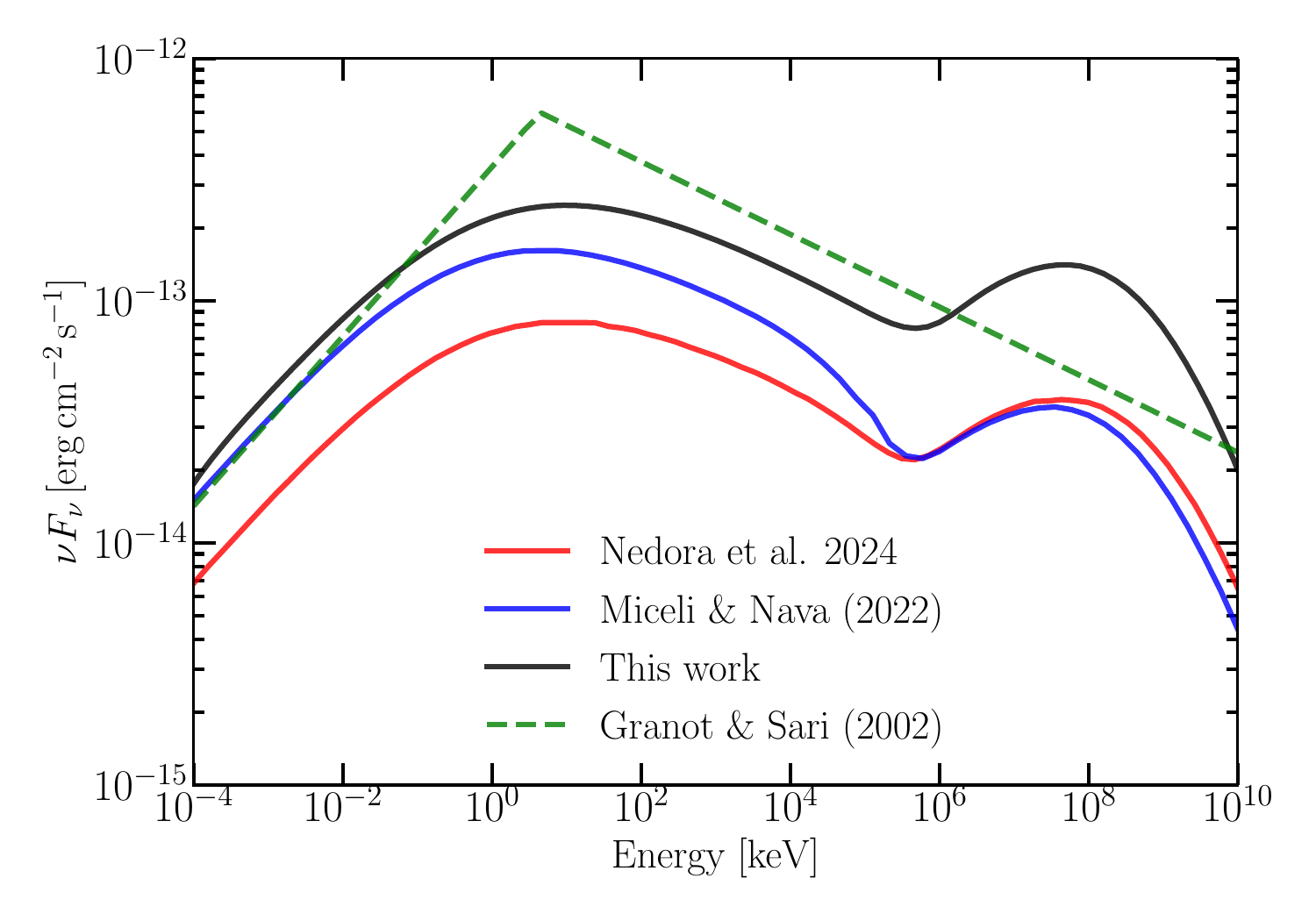}
    \includegraphics[width=0.33\textwidth]{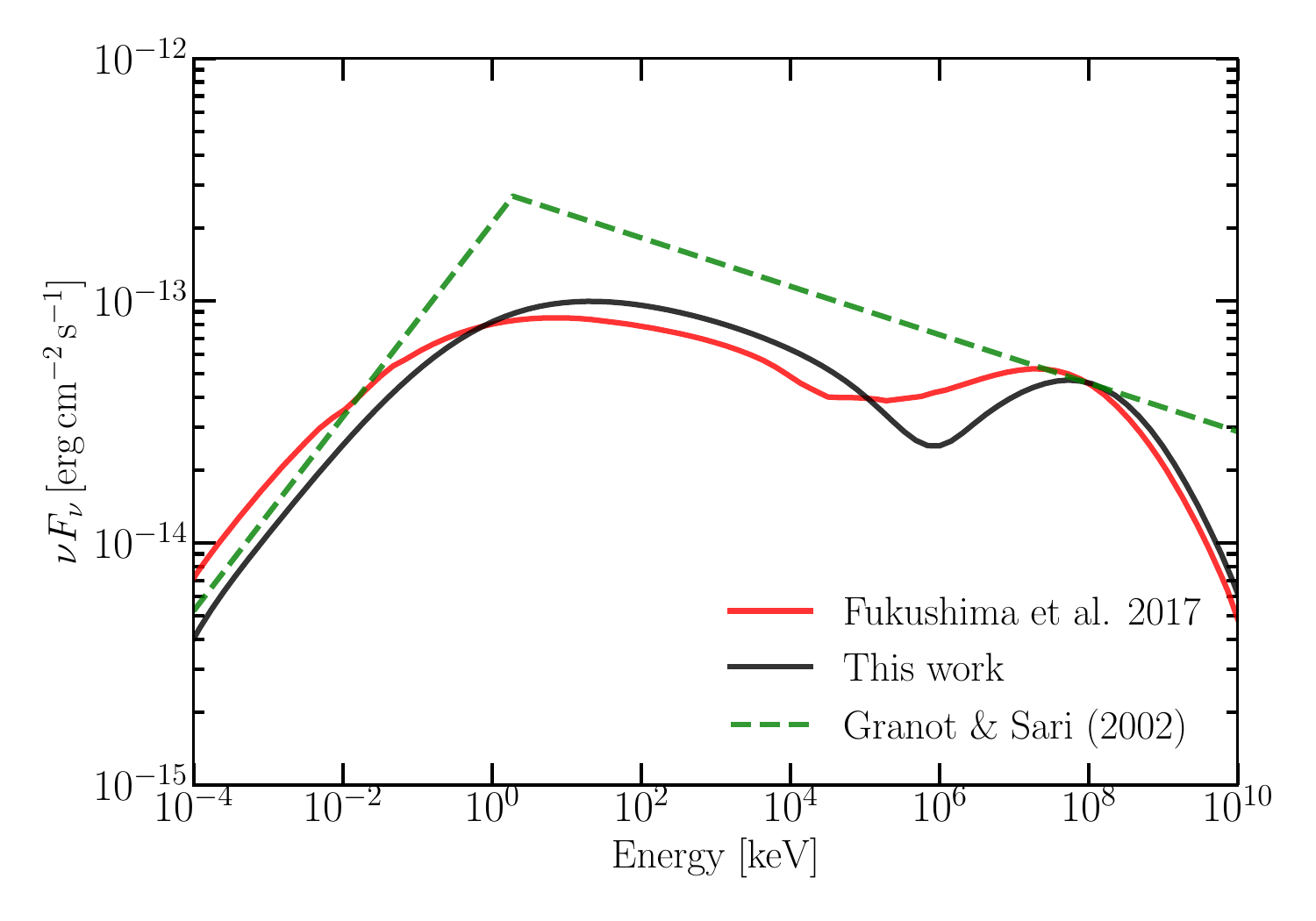}
    \includegraphics[width=0.33\textwidth]{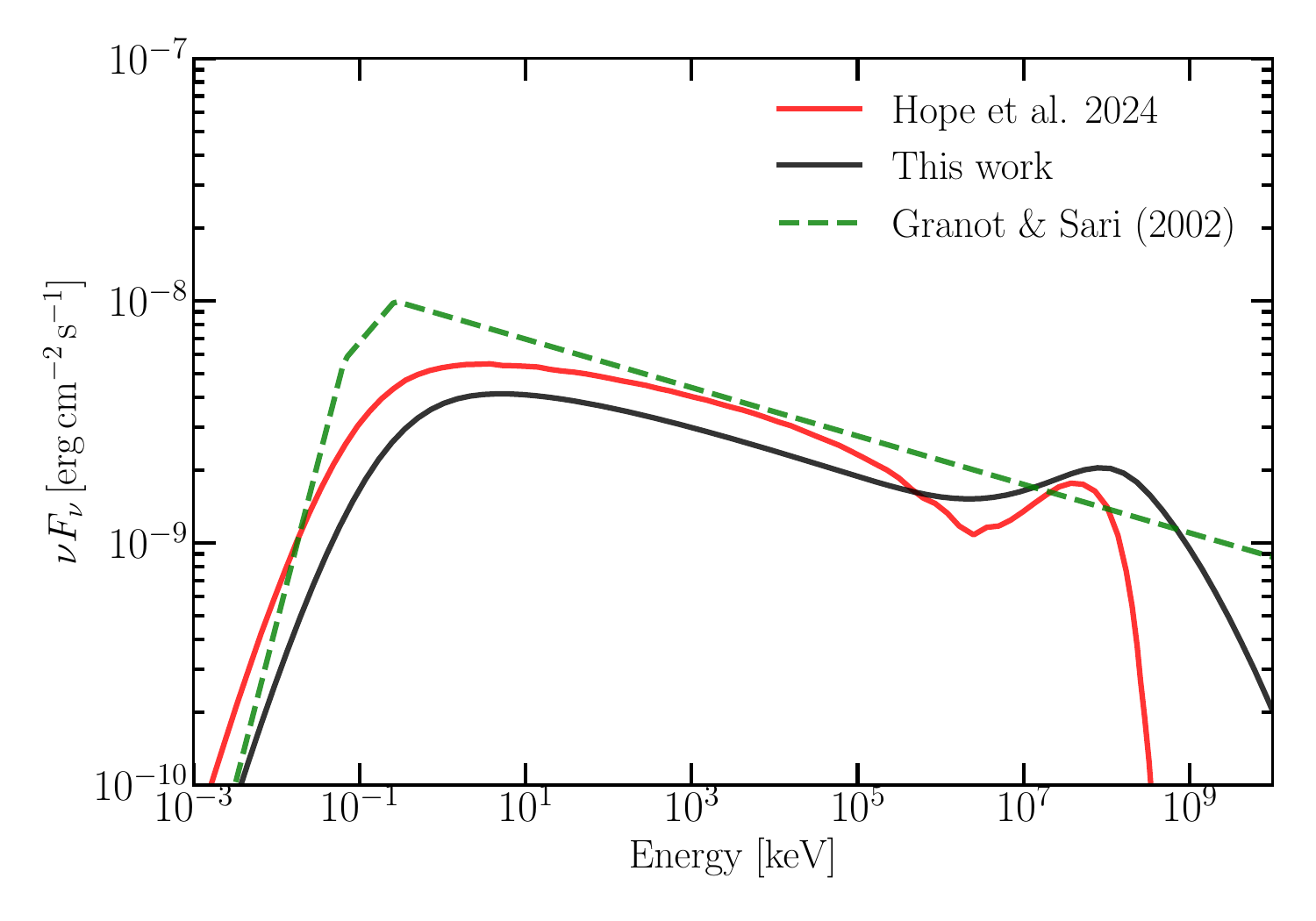}
    \caption{
    (\textbf{Left}) Spectral comparison with \citet{Nedora+24} and \citet{Miceli-Nava-22}. We do not apply the $\gamma\gamma$ supression at TeV energies due to the EBL, while the other two works do. The work and that of \citet{Nedora+24} perform integration over the EATS, but only LOS spectra is given in \citet{Miceli-Nava-22}. The model parameters assumed here are: $E_{\rm k,iso}=2\times10^{52}\,{\rm erg}$, $\Gamma_0=400$, $n_{\rm ISM} = 1\,{\rm cm}^{-3}$, $p=2.3$, $\epsilon_e = 5\times10^{-2}$, $\epsilon_B=5\times10^{-4}$, $\xi_e=1$, with $z=1$ which corresponds to $d_L=2.1\times10^{28}$\,cm, and $t_{\rm obs}=10^4$\,s. 
    (\textbf{Middle}) Spectral comparison with \citet{Fukushima+17}. Both works perform integration over the EATS and do not apply the $\gamma\gamma$ supression at TeV energies due to the EBL. The model parameters assumed here are: $E_{\rm k,iso}=10^{52}\,{\rm erg}$, $\Gamma_0=100$, $n_{\rm ISM} = 1\,{\rm cm}^{-3}$, $p=2.2$, $\epsilon_e = 0.1$, $\epsilon_B=10^{-3}$, $\xi_e=1$, with $z=2$ which corresponds to $d_L=4.92\times10^{28}$\,cm, and $t_{\rm obs}=10^4$\,s. 
    (\textbf{Right}) Spectral comparison with \citet{Hope+25}. Both works perform integration over the EATS but only \citet{Hope+25} apply the $\gamma\gamma$ supression at TeV energies due to the EBL. The model parameters assumed here are: $E_{\rm k,iso}=10^{53}\,{\rm erg}$, $\Gamma_0=2\times10^3$, $n_{\rm ISM} = 1\,{\rm cm}^{-3}$, $p=2.2$, $\epsilon_e = 0.1$, $\epsilon_B=10^{-2}$, $\xi_e=1$, with $z=0.5454$ which corresponds to $d_L=10^{28}$\,cm, and $t_{\rm obs}=10^2$\,s.
    } 
    \label{fig:spec_comparison}
\end{figure*}

\subsection{Numerical Scheme and Grids}
We solve the coupled kinetic equations for particles and photons over a logarithmic grid of dimensionless momenta ($p\equiv\gamma\beta_e$) and photon energy, both with $N_{\rm Grid}=320$ points. The momentum grid spans over eleven orders of magnitude with $10^{-3}\leq p\leq10^8$ while the photon energy grid spans fifteen orders of magnitude with $10^{-8}\leq x\leq10^7$. The kinetic equations are solved using the \citet{Chang-Cooper-70} implicit finite difference scheme, which is only accurate to first-order in both space and time but it offers more stability than any of the higher order schemes \citep[e.g.][]{Park-Petrosian-96} due to it being unconditionally stable. In addition, as also true for implicit schemes in general, this scheme is free of the Courant condition where the size of the timestep does not depend on the size of the energy grid; the contrary is always true for explicit numerical schemes. However, even here care should be taken, and to ensure convergence to the accurate solution the step size is limited by that determined by the fastest cooling time \citep[see, e.g., Fig.\,2 in][]{Aguilar-Ruiz+26}.

\subsection{Photon Escape \& EATS Integration}
We adopt the treatment of \citet{Lightman-Zdziarski-87} to describe the escape of radiation from the dissipation zone, 
with photons escaping at a rate prescribed by $\dot{n}'_{\gamma,\rm esc}(x') = -n'_\gamma(x')/t'_{\rm esc}(x')$, and 
over an energy-dependent escape time $t'_{\rm esc}(x')$,
\begin{equation}
t'_{\rm esc}(x') = \frac{2\Delta'}{c}\left[1+\frac{1}{3}f(x')\tau_{\rm KN}(x')\right]\,,
\end{equation}
where the factor of 2 signifies escape from the front and back of the emission that is locally treated as a slab \citep[e.g.][]{Fukushima+17}. The light-crossing time from the emission region is modified by energy-dependent electron scattering. Here $\tau_{\rm KN}(x')$ is the scattering optical depth of electrons obtained using the full Klein-Nishina scattering cross-section, and the factor
\begin{equation}
f(x') = \left\{\begin{array}{ll}
    1~, & x'\leq 0.1 \\
    (1-x')/0.9~, & 0.1<x'<1 \\
    0~, & x' \geq 1
\end{array}\right.
\end{equation}
gives a smooth transition from non-relativistic to relativistic scattering, in which case forward scattering 
becomes important. However, in most cases the flow remains optically thin to scattering and therefore the 
scattering term in the escape time can be ignored, and we only include it here for completeness.

Over a time-step $\Delta t'$, the \textit{isotropic} radiation field that escapes is given by
\begin{eqnarray}
    \Delta n_{\gamma,\rm esc}'(x',t') &&= n_\gamma'(x',t'-\Delta t') - n_\gamma'(x',t') \\
    &&= n_\gamma'(x',t'-\Delta t')\left[1-\exp\fracb{-\Delta t'}{t'_{\rm esc}(x')}\right] \,.
\end{eqnarray}
From this we can obtain the isotropic emissivity
\begin{equation}
    j'_{\nu',\rm iso}(x',t') = hx'\frac{\Delta n'_{\gamma,\rm esc}(x',t')}{\Delta t'} = \frac{L'_{\nu',\rm iso}}{V'}
\end{equation}
where $L'_{\nu',\rm iso}$ is the spectral luminosity. We save the isotropic emissivity at every time-step in the comoving frame, and then use it to obtain the observed flux density by integrating over the equal arrival time surface (EATS).

At any given observer-frame time $t_{\rm obs}$ photons arrive from different angles away from the 
line-of-sight (LOS) that were emitted at different lab-frame times $t$ or corresponding radial distance 
$R$ of the emitting shell from the central engine. Therefore, even though the observed flux is 
dominated by that emitted close to the LOS, an accurate treatment demands integration over the EATS. 
This effect is governed by the following equation
\begin{equation}
    t_{\rm obs,z} = \frac{t_{\rm obs}}{1+z} = t - \frac{R\tilde\mu}{c}\,,
\end{equation}
where $\tilde\mu = \hat R\cdot\hat n = \cos\tilde\theta$. With the coordinate system origin at the 
location of the central engine, $\hat R$ is the radial unit vector of the 
material from which emission is received, $\hat n$ is the unit vector along the direction of the observer, 
and $\tilde\theta$ is the polar angle measured from the LOS to the emitting material. The lab-frame time is obtained from the 
dynamics of the thin-shell,
\begin{equation}
    t = \frac{R_d}{c}\int_0^\chi \frac{d\chi'}{\beta(\chi')}\,,
\end{equation}
where $\beta(\chi) = \sqrt{1-\Gamma^{-2}(\chi)}$. The flux density observed from a source at redshift 
$z$, with a corresponding luminosity distance $d_L(z)$, is then obtained from \citep[e.g.][]{Gill-Granot-18a}
\begin{equation}
    F_{\nu}(\nu,t_{\rm obs}) = \frac{(1+z)}{16\pi^2d_L^2}\int\delta_D^3L_{\nu',\rm iso}'(\nu',t')d\Omega\,,
\end{equation}
where $\delta_D = [\Gamma(1-\beta\tilde\mu)]^{-1}$ is the Doppler factor and $d\Omega=d\tilde\mu\,d\tilde\varphi$ 
is the unit solid angle subtended by a fluid element on the expanding shell onto the central engine, with 
$\tilde\varphi$ being the azimuthal angle measured around the LOS. The observed frequency $\nu$ is related 
to the comoving one via the Doppler boost and redshift, such that $\nu = (1+z)^{-1}\delta_D\nu'$. The 
comoving time is obtained from the lab-frame time for a given $t_{\rm obs}$ and $\tilde\mu$ via the 
time-dilation, with $t'=t/\Gamma$.

\subsection{Comparison with Other Works}
In this section we present spectral comparisons with other works that employ similar numerical methods and formalism to calculate the high-energy afterglow emission. All of these different works make slightly different assumptions in calculating the blast wave dynamics, the treatment of the comoving emission and its escape, and transfer of radiation from the comoving to the observer frame. Therefore, small differences in the results are expected.

In the left panel of Figure\,\ref{fig:spec_comparison}, we compare the spectrum from our code to that from \citet{Nedora+24} and \citet{Miceli-Nava-22} for the same model parameters. We find that the spectral shape obtained in this work and in \citet{Nedora+24} are most similar, but the normalization differs by a factor $\sim2$. Both results perform an integration over the EATS. In contrast, the work of \citet{Miceli-Nava-22} does not perform EATS integration and simply give the spectrum obtained from radiation received along the LOS, which is also the most common assumption made by many analytical works \citep[e.g.][]{Sari+98}. Finally, we compare our results with the widely used analytical model of the synchrotron afterglow of \citet{Granot-Sari-02} that not only performs an integration over the EATS but also over the entire shocked volume behind the forward shock. The results from our code, in terms of the spectral peak and its normalization, are in excellent agreement with that of \citet{Granot-Sari-02}, who do not calculate the SSC component or its effect on the seed synchrotron spectrum.

Both \citet{Nedora+24} and \citet{Miceli-Nava-22} apply the suppression at very-high-energies due to $\gamma\gamma$ absorption of TeV photons on the much softer photons of the extra-galactic background light (EBL), which is mostly composed of the cosmic microwave background and stellar ultraviolet photons from galaxies. We do not apply such correction since it is model dependent and different works use different models. On the other hand, we do include internal absorption of high-energy photons due to $\gamma\gamma$-annihilation. For afterglow shocks, the optical depth is much smaller than unity for most of the spectrum except around a TeV and above where a small suppression may appear.

The middle panel of Figure\,\ref{fig:spec_comparison} compares our results with that from \citet{Fukushima+17}. There is generally good agreement between the two results and both also agree well with the spectrum from \citet{Granot-Sari-02}. Finally, the right panel makes a comparison with the recent work of \citet{Hope+25}, where we find a slight difference in the normalization. \citet{Hope+25} also apply the suppression at high energies due to the EBL and therefore the flux drops very rapidly. There is generally good agreement between the two spectra for the synchrotron component, however, we find a much broader inverse-Compton peak.

\section{Neural Emulation and MCMC inference}\label{sec:NN&MCMC}

\subsection{Grid generation}\label{subsec:gridgen}

Our model has in total six parameters that includes the isotropic-equivalent kinetic energy ($E_{\rm k,iso}$), coasting Lorentz factor ($\Gamma_0$), number density of the ISM ($n$), the shock microphysical parameters ($\epsilon_e$ and $\epsilon_B$), and the electron energy distribution power-law index ($p$). In this work, we specifically consider an ISM environment, for which $k=0$, and also fix the electron acceleration fraction to $\xi_e=1$. To train our ANN surrogate, we construct the model grid over these six parameters as well as over discretized observer-frame time $t_{\rm obs}$. Therefore, in total, there are seven input parameters for our neural network. We obtain the spectra in the cosmological rest frame of the GRB (i.e. $z=0$) while placing the source at a luminosity distance of $d_L=10^{28}$\,cm. To obtain the spectrum for $z>0$, we translate it in frequency space by using $\nu_* = \nu (1+z)$ and appropriately adjusting the flux density normalization so that $F_\nu(\nu,t_{\rm obs})=(1+z)[10^{28}\,{\rm cm}/d_L]^2F_{\nu_*}(\nu_*,t_{\rm obs*})$ with $t_{\rm obs*}=t_{\rm obs}/(1+z)$, where starred quantities are obtained in the GRB rest frame.

We prepare a discretized grid over the model parameters, as shown in Table\,\ref{tab:model_grid}. The ranges of the different parameters were chosen based on afterglow fits obtained in the literature. For example, the value of $\epsilon_e$ has typically been found to be around $0.1$, with most inferred values lying between $\sim0.02$ and $0.6$ \citep{Santana+14}. The equipartition value of $\epsilon_e=1/3$ is obtained when allowance is made for equal energy to be shared between protons, electrons, and the magnetic field. In comparison, the allowed range for $\epsilon_B$ is much larger, with inferred values generally ranging between $10^{-8}$ and $10^{-3}$ \citep{Santana+14,Miceli-Nava-22}. A caveat to note here is that many analytical works do not perform multi-wavelength afterglow fits to determine the shock-microphysical parameters accurately, and typically fix other model parameters to constrain them. Typical values ranging from $10^{52}$\,erg to $10^{54}$\,erg have been obtained for $E_{\rm k,iso}$ from broadband afterglow fits \citep{Meszaros-06,Kumar-Zhang-15}, while values as high as $10^{55}$ erg were required for modeling energetic events such as the bright GRB 221009A \citep{OConnor+23,Gill-Granot-23}. The electron energy distribution power-law index $p$ can be inferred from the synchrotron spectrum, and it commonly ranges from $2$ to $3$, with many observations being compatible with values between $2.2$ and $2.5$ \citep{Meszaros-06}. The ISM density has a large spread and it can vary between $10^{-3}\,{\rm cm}^{-3}$ to $100\,{\rm cm}^{-3}$ \citep{Soderberg+06}, with larger densities found in the environments of collapsar-driven GRBs and lower densities in merger-driven GRBs. The coasting Lorentz factor of GRB jets is limited from below, with typical values around $\Gamma\sim100$ that avoids the compactness problem \citep{Piran-04}. Larger values up to $\Gamma_0\sim10^3$ have been inferred in some GRBs that showed high-energy spectral cutoffs due to $\gamma\gamma$-annihilation or $\sim$GeV energy photons in the prompt emission spectrum \citep[e.g.][]{Abdo+09,Tang+15}.

Guided by these literature values, in this work we adopt the following parameter domain for the training grid. We use $2.01 \leq p \leq 2.6$, $10^{-1} \leq n_0 \leq 10$ cm$^{-3}$, and $100 \leq \Gamma_0 \leq 600$, together with the ranges of $E_{\rm k,iso}$, $\epsilon_e$, and $\epsilon_B$ listed in Table \ref{tab:model_grid}. Finally, we construct a logarithmic grid for the observer-frame time with a range $1\leq \log(t_{\rm obs*}\,[{\rm s}])\leq5$ over 12 grid points that includes the two limiting values. We find the temporal grid is broad enough to cover any TeV emission due to SSC at early times. Using the parameter values from Table \ref{tab:model_grid}, we prepare a grid of spectra, i.e. flux density in mJy as a function of frequency in Hz, using the radiation code described in Section \ref{sec:radiation_code} over a wide logarithmic frequency grid with $10\leq\log\nu_*\,({\rm Hz})\leq30$. The total number of points in the grid is given by the product of the number of all model parameter values and the temporal grid points. Therefore, the final grid consists of 100,800 model spectra spanning the selected parameter space. 

Our observer-frame spectral grid, with frequency $\nu_*$, is much wider than our comoving grid that has the frequency range of $1.2\times10^{12}\leq\nu'\,(\rm{Hz})\leq1.2\times10^{27}$. For radiation received along the line-of-sight the Doppler boosted photon frequency is $\nu_* = 2\Gamma\nu'$, which will shift the lowest energy comoving grid point to higher energies by at most a factor of $200 \leq 2\Gamma \leq 1200$ depending on the $\Gamma$ value. As a result, no flux value is recorded at $\nu_*<2\Gamma\nu'_{\min}$, which presents numerical problems for the ANN as it has no information there to build the spectrum for prediction. To circumvent this issue, we manually extrapolate the spectrum to lower energies based on the local spectral slope from the available spectrum at low energies. A similar extrapolation is performed near the higher grid boundary.

\begin{table}
\centering
\caption{
Varying parameters in the model grid used to train the neural network. We use fixed parameters: $\xi_e = 1$, $z = 0$, $d_L = 10^{28} \rm \, cm$ an $k = 0$ (for ISM). The last column indicates the number of values used in the grid for each parameter.}
\begin{tabular}{l l c}
\hline
Parameter & Values & $N$\\
\hline
$\log E_{\rm k,iso}$ [erg] & $52, \, 53, \, 54, \, 55, \, 56 $ & 5 \\
$\log \Gamma_0$ & $\log(100), \log(300), \log(600)$ & 3\\
$\log n_0$ [cm$^{-3}$] & $-2, \, -1, \, 0, \, 1$ & 4\\
$\log\epsilon_e$ & $-3, \, -2.5, \, -2, \,-1.5, \, -1$ & 5\\
$\log \epsilon_B$ & $-5, \, -4.5, \, -4, \, -3.5, \, -3, \, -2.5, \, -2$ & 7\\
$p$ & $2.01, 2.2, 2.4, 2.6$& 4 \\
\hline
\end{tabular}
\label{tab:model_grid}
\end{table}

\subsection{Artificial Neural Network}

\subsubsection{Setup and training}

The model emulation of GRB afterglows using ANNs is a complex problem, particularly due to the degenerate and highly non-linear nature of the underlying physical processes. To efficiently capture the relationship between the physical parameters and the resulting spectra, we adopt a decoupled approach by splitting the training task into two components: spectral shape and flux amplitude. We implemented this surrogate model using the Python framework \texttt{PyTorch}\footnote{https://pytorch.org} \citep{Paszke+19}, an open-source deep learning framework favored for its flexibility, efficient tensor operations and GPU acceleration capabilities. 

We decompose the spectral prediction into two targets: the normalized spectral shape and the overall flux normalization. A Deep Residual Network \citep[ResNet;][]{He+16} maps the seven input parameters onto the normalized spectral shape, while a multilayer perceptron (MLP) independently predicts the scalar flux normalization. The residual architecture was adopted for the higher-dimensional spectral-shape output, whereas the simpler MLP is used for the scalar normalization. Separating these two targets allows the networks to learn variations in spectral shape independently of the overall flux scale.

For the spectral shape emulation, the ANN consists of four residual blocks, each one with a width of 512 neurons. For this network we adopted an initial learning rate of $8\times10^{-4}$ and implemented regularization by setting a weight decay of $1 \times 10^{-4}$ and a dropout rate of 5\%. 
During training, dropout randomly sets a fraction of the network activations to zero at each step.
These regularization choices are intended to reduce overfitting and improve predictive performance on spectra not used during training.
For the scalar flux normalization, we use an MLP network with four layers of 256 neurons each. For this network, the regularization is set by an initial learning rate of $1 \times 10^{-3}$ and weight decay of $1 \times 10^{-4}$. 

Training was performed on 80\% of the total grid, with the remaining 20\% reserved for validation. Both input parameters and output targets were standardized by removing the mean and scaling to unit variance. We used the \texttt{AdamW} optimizer for both networks, which implements weight decay separately from the gradient-based parameter update. We found that this choice significantly reduced the RE compared with the \texttt{Adam} optimizer. The training is driven by the minimization of a loss function given by mean squared error (MSE) \footnote{${\rm MSE} = \frac{1}{N} \sum\limits_{i=1}^N (y_{i, {\rm true}} - y_{{i, \rm pred}})^2$ where $y_{\rm true}$ and $y_{\rm pred}$ are the actual and predicted values in logarithm space.}, which quantifies the discrepancy between the predicted and the true values. Because the MSE depends quadratically on the residuals, larger prediction errors contribute more strongly to the loss. We use the Gaussian Error Linear Unit (GELU) activation function in all layers, as its smooth activation is suitable for the smooth transitions and curvatures inherent in the physical model. Training was performed for 450 epochs. The learning rate was adjusted during training using a cosine-annealing schedule. Finally, during training we set a batch size of 512. We use a single 80/20 training-validation split and do not perform cross-validation. Consequently, the ``on-grid'' (i.e. using spectra produced for input parameters that form the nodal points of the $N$-dimensional model grid) validation results reported below characterize the predictive performance for this particular split and do not quantify its sensitivity to the choice of training and validation samples. However, we also verified the robustness of these results by repeating the training for five independent training-validation splits. The complete setup and hyperparameter values are summarized in Table \ref{tab:ANN_setup}.

\begin{figure}
    \centering
    \includegraphics[width=\linewidth]{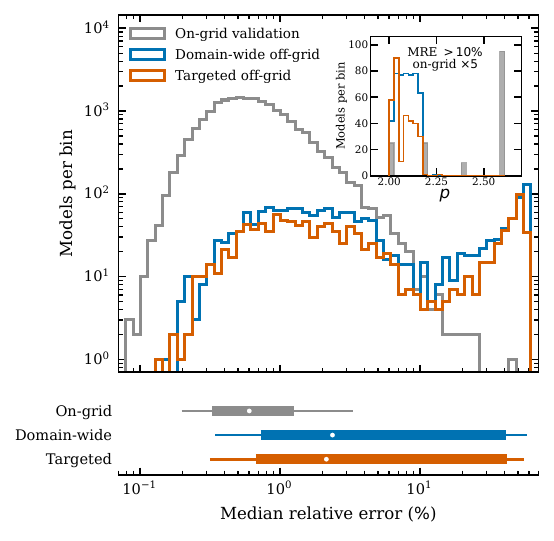}
    \vspace{-19pt} 
    \caption{Distribution of the median relative errors (MREs) for the on-grid validation, domain-wide off-grid, and targeted off-grid samples defined in Section \ref{subsec:modelvalidation}. The upper panel shows the number of models in each MRE bin. The lower panel shows the median (circles) and central 68 and 95 per cent intervals (thick and thin lines, respectively). The inset shows the distribution of $p$ for models with MRE $>10$\,per cent; the on-grid counts are multiplied by five for visibility.
    }

    \label{fig:MRE_distribution}
\end{figure}

\subsubsection{Model Validation}\label{subsec:modelvalidation}

We assess the surrogate model using the magnitude of the relative error (RE) between the ANN predictions and the corresponding kinetic-model spectra. Because the RE varies with frequency, we characterize the accuracy of each predicted spectrum by the median RE over the full frequency domain, which we denote as the median relative error (MRE).
We consider three evaluation samples. In addition to the on-grid validation sample, which contains 20\,160 spectra from the 20 per cent of the original Cartesian model grid withheld during training, we test interpolation between grid nodes. For this purpose, we calculated a further 3000 kinetic-model spectra at parameter combinations lying strictly off the original grid (``off-grid''; i.e. spectra produced using input parameters that are in between the nodal grid points but still within the global boundary of the $N$-dimensional grid). Of these, 1800 are generated using a scrambled seven-dimensional Sobol sequence spanning the full parameter domain. This ``domain-wide" sample provides approximately uniform, space-filling coverage in the normalized input coordinates. The remaining 1200 spectra deliberately target regions expected to be more difficult to interpolate. This ``targeted" sample contains 450 points near individual boundary surfaces, 300 near intersections of two boundaries, 300 near intersections of three or four boundaries, and 150 within the largest 10 per cent of the original grid cells by normalized hypervolume. For the boundary subsets, the selected coordinates lie within 1-5 per cent of either boundary, while the remaining coordinates lie within the central 10-90 per cent of their allowed ranges. Thus, the domain-wide sample tests interpolation throughout the parameter space, whereas the targeted sample preferentially probes boundaries and the most sparsely sampled regions of the original grid.

Fig. \ref{fig:MRE_distribution} compares the MRE distributions of the three samples. The on-grid validation distribution is strongly concentrated at low MRE, with a median of 0.6 per cent. Its central 68 per cent (95 per cent) interval spans 0.3--1.3 per cent (0.2--3.3 per cent). Of the 20\,160 spectra in this sample, 31 ($\approx 0.15$\, per cent) have MREs above 10 per cent. The domain-wide sample has a median MRE of about 2.4 per cent, with central 68 per cent (95 per cent) intervals of 0.7--41 per cent (0.35--58 per cent). The targeted sample has a similar distribution (median of about 2.1 per cent, 68 and 95 per cent intervals of 0.7--42 per cent and 0.3--55 per cent). In total, 495 of the 1800 domain-wide spectra (27.5 per cent) and 319 of the 1200 targeted spectra (27 per cent) have MREs above 10 per cent.

\begin{figure}
    \centering
    \includegraphics[width=\linewidth]{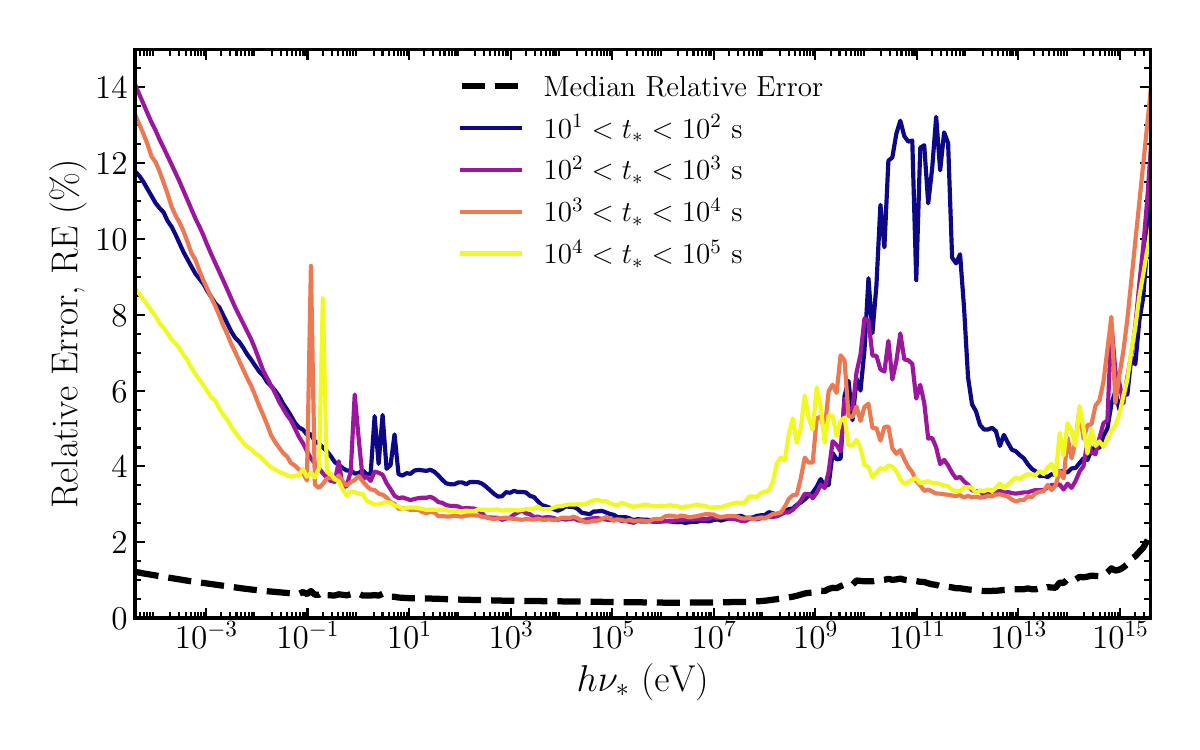}
    \vspace{-19pt} 
    \caption{Relative error as a function of source-frame photon energy in the on-grid validation sample. Solid lines show the 97.5th percentile of the relative error for each time decade in the on-grid validation sample, while the dotted line represents the 50th percentile (median) of the relative error for all time decades.
    } 
    \label{fig:RE_vs_energy_per_decade}
\end{figure}

The off-grid high-error population is strongly localized in the electron energy distribution power-law index $p$. Of the 814 off-grid spectra with MREs above 10 per cent, 812 have $p<2.2$, predominantly within $2.01<p\lesssim 2.18$. By contrast, only two of the 2057 off-grid spectra with $p\geq 2.2$ exceed this threshold. The on-grid validation sample contains only the discrete grid values $p=2.01, 2.2, 2.4,$ and 2.6 and therefore does not sample the affected intermediate values. No comparable systematic dependence on the remaining input parameters is found. Both applications of the surrogate model presented in this paper lie outside the affected region: the synthetic test uses $p=2.4$, while the fit to GRB 190114C gives $p=2.31^{+0.04}_{-0.04}$. The identified limitation is therefore not expected to affect the results presented here. Denser sampling near $p=2$ will be investigated in future extensions of the model grid.

The results shown in Fig. \ref{fig:RE_vs_energy_per_decade} indicate excellent overall performance, with the overall median RE (dashed line in figure) over all times ranging from $\sim 0.5$ to $2.0 \%$ across the entire frequency domain. To assess the performance of our ANN we examine the 97.5th percentiles for different time intervals, which represents the largest possible error at any given frequency but not the most likely. These results demonstrate the robustness of our surrogate model, which maintains high precision with RE values below $\sim 7\%$, particularly within the energy range from $10^{-1} \, \rm eV$ to $10^{14} \, \rm eV$. However, at early times ( $10^{1}-10^{2} \, \rm s$ ) and in the energy band of $\sim 10^{10}-10^{12}$ ~eV, the 97.5th percentile exhibits larger RE values,  exceeding 10\%. For TeV GRB afterglows, the higher RE observed at early times ($10^{1}-10^{2}$ s) within the 0.1–1 TeV band is in fact typically smaller than the observational uncertainties in the GRB 190114C data analyzed in this work (as described in Section \ref{subsec:MCMC}, the mean relative observational uncertainty above 10 GeV is approximately 30\%).

Furthermore, the 97.5th percentiles indicate that the error increases at the highest energies for all time intervals, reaching up to 14\% at $1 \, \rm PeV$. A similar trend is observed below $\sim 10^{-1} \, \rm eV$ at all times, where the largest errors occur in the $10^{2}-10^{3}$ s interval, while smaller errors are found in the $10^{4}-10^{5}$ s interval. This increase in RE is due to the ANN not performing so well near the spectral grid boundaries as expected. However, the large errors above $100 \, \rm TeV$ (in the GRB rest frame) does not impact GRB afterglow modeling for source redshift $z\lesssim5$ as the most energetic photons detected to date reach only $E_\gamma \sim 18 \, {\rm TeV} /(1+z)$ \citep[i.e.,][]{Huang+2022}. Since TeV emission can only be observed to much lower redshifts due to $\gamma\gamma$-absorption on EBL photons, these large errors will never impact the fit results. On the other hand, the larger errors below $10^{-1}\, \rm eV$ could impact model comparison with radio to IR observations, as they may exceed the characteristically small observational uncertainties in this energy range.  We therefore consider $0.1 \, \rm eV$--$100 \,\rm TeV$ to be the most reliable energy range of our model.

Our results confirm that the surrogate model is accurate on average across the on-grid validation sample, reproducing the synchrotron and SSC components over more than 15 orders of magnitude in energy.

\begin{table}
\centering
\caption{
Summary of the hyperparameters, architecture and training configuration of the artificial neural network used in this work.
The model consists of two independent networks: a deep residual network  (ResNet) for predicting the spectral shape and a direct multilayer perceptron network (MLP) for predicting the logarithmic amplitude. }
\begin{tabular}{l l }
\hline

\multicolumn{2}{c}{\textbf{General Setup}} \\
\hline
Inputs & $\log E_{\rm k,iso}, \, \log\Gamma_0, \, \log n_0, \, \log\epsilon_e, \, \log \epsilon_B, \, p, $
                         \\ & $  \, \log t_{\rm obs*} $ \\
Training Batch size & 512 \\
Training Epochs     & 450 \\
Activation function & GELU (All layers)\\
Loss function       & MSE (log-space) \\
Optimizer algorithm     & AdamW \\
Data scaling        & Standardization \\
Training fraction & 0.8 \\
\hline

\multicolumn{2}{c}{\textbf{Shape Network}} \\
\hline
Deep Residual& 4 blocks (width = 512) \\
Dropout         & 0.05 \\
Initial Learning rate   & $8 \times 10^{-4}$ \\
Weight decay            & $1 \times 10^{-4}$ \\
Target                  & Normalized Spectral prediction $\log( \nu_* F_{\nu_*} / F_{\max})$ \\
\hline

\multicolumn{2}{c}{\textbf{Amplitude Network}} \\
\hline
Multilayer Perceptron   & (256, 256, 256, 256) \\
Initial Learning rate  &  $1 \times 10^{-3}$ \\
Weight decay           &  $1 \times 10^{-4}$ \\
Target                 & Scaling factor $\log(F_{\max})$ \\
\hline

\multicolumn{2}{c}{\textbf{Training Strategy}} \\
\hline
Learning rate schedule & Decay strategy by Cosine annealing \\
Model structure        & Separate shape and amplitude networks \\
\hline
\end{tabular}
\label{tab:ANN_setup}
\end{table}

\subsection{Markov Chain Monte Carlo Inference}\label{subsec:MCMC}

To infer the physical parameters from TeV GRB observations, we perform likelihood-based Bayesian inference using Markov Chain Monte Carlo (MCMC) sampling, implemented using the publicly available Python package \texttt{emcee}\footnote{https://emcee.readthedocs.io} \citep{Foreman-Mackey+13}. This sampler is well suited for efficiently sampling high-dimensional parameter spaces with strong correlations. The agreement between the fluxes predicted by our SSC model and the observed multi-epoch, multi-frequency data is quantified by the likelihood function. Assuming Gaussian errors, the (log-)likelihood is defined as
\begin{equation}\label{eq_loglikelihood}
\log \mathcal{L_{\rm total}} = -\frac{1}{2} \sum_i \left[ 
\frac{ \left([\nu F_\nu]_{i}^{\rm obs} - [\nu F_\nu]_{i}^{\rm model} \right)^2 }{ { \sigma_{i}^2} } 
+ \ln(2\pi \sigma_{i}^2) 
\right],
\end{equation}
where $[\nu F_\nu]_{i}^{\rm obs}$ and $[\nu F_\nu]_{i}^{\rm model}$ denote the observed and modeled fluxes, respectively, for the $i$-th observation. The total uncertainty $\sigma_{i}$ associated with each point is the sum under quadrature of three terms: 
\begin{eqnarray}
    \sigma_i^2 =\sigma_{{\rm obs}, i}^2 + \sigma_{{\rm NN}, i}^2 + \sigma_{{\rm intr}, i}^2 \, ,
\end{eqnarray}
where $\sigma_{{\rm obs}, i}$ is the reported observational uncertainty, $\sigma_{{\rm NN},i}$ accounts for the finite accuracy of the neural-network surrogate, and an intrinsic scatter term $\sigma_{{\rm intr},i}$ that captures residual variance not accounted for by the first two terms. 

In this work, we assume zero additional intrinsic scatter ($\sigma_{{\rm intr},i} = 0$) as its usage is more appropriate when using phenomenological models rather than physical models, which in our case, have a well-defined prediction with no intrinsic uncertainty. We account for the finite accuracy of the ANN surrogate through the frequency- and time-dependent term $\sigma_{{\rm NN},i} = {\rm RE}_{97.5} \, [\nu F_\nu]_{i}^{\rm model}$, where ${\rm RE}_{97.5}$ represents the 97.5th percentile of the surrogate relative error distribution in the on-grid validation sample (see Fig.~\ref{fig:RE_vs_energy_per_decade}). The posterior distribution is obtained by combining the likelihood (Eq.~\ref{eq_loglikelihood}) with uniform priors defined over values typically inferred from GRB afterglow observations and theoretical considerations.  To assess chain convergence and ensure reliable posterior inferences, we compute the integrated autocorrelation time $\tau_i$ for each parameter $i$. A chain is considered sufficiently long when its length per walker $N_{\rm steps}$ satisfies $N_{\rm steps} \ge 50 \max(\tau_i)$ \citep{Foreman-Mackey+13}. For post-processing, we discard the initial burn-in phase corresponding to $N_{\rm discard} = \lceil 2 \max(\tau_i) \rceil$ steps per walker to eliminate dependence on initial conditions. To yield statistically independent posterior samples and reduce memory overhead, the remaining chain is thinned by taking every $k$-th sample, where $k = \max(1, \lfloor 0.5 \min(\tau_i) \rfloor)$. Final parameter estimates and uncertainties are reported using the median ($50\text{th}$ percentile) and $16\text{th}$--$84\text{th}$ percentile credible intervals, respectively.

We tested our method by fitting synthetic data generated by the kinetic code, using the following parameters: 
$\epsilon_e = 10^{-2}$, $\epsilon_B = 10^{-4}$, $E_{k, \rm iso} = 10^{54} \, \rm erg$, $\Gamma_0 = 400$, $p = 2.4$, and $n = 1 \, \rm cm^{-3}$. 
In addition, we introduce synthetic observational uncertainties to emulate real observing conditions. We estimate representative relative observational uncertainties, $\sigma_{{\rm obs}, i}/[\nu F_\nu]_{i}$, by averaging the uncertainties of the GRB 190114C data within three energy ranges. We adopt fractional uncertainties of 7\% below 1 MeV, 66\% between 1 MeV and 10 GeV, and 30\% above 10 GeV. These values also provide a reference for assessing the surrogate errors discussed in Section \ref{subsec:modelvalidation}. Figure \ref{fig_CORNER_synthetic} shows that the injected parameter values lie within the $1\sigma$ credible intervals of the inferred posterior distributions, demonstrating successful parameter recovery for this synthetic test.
The deviations between the true values and the median inferred values are quantified as relative errors in Table \ref{tab:relative_errors}. 
The errors range from 0.5\% for $\epsilon_B$ to 9.9\% for $E_{k, \rm iso}$.

\section{Model Fit to the Afterglow of GRB\,190114C}\label{sec:model_fits}

\begin{figure*}
\centering
\begin{minipage}[b]{0.50\linewidth}
\centering
\includegraphics[width=1.\linewidth]{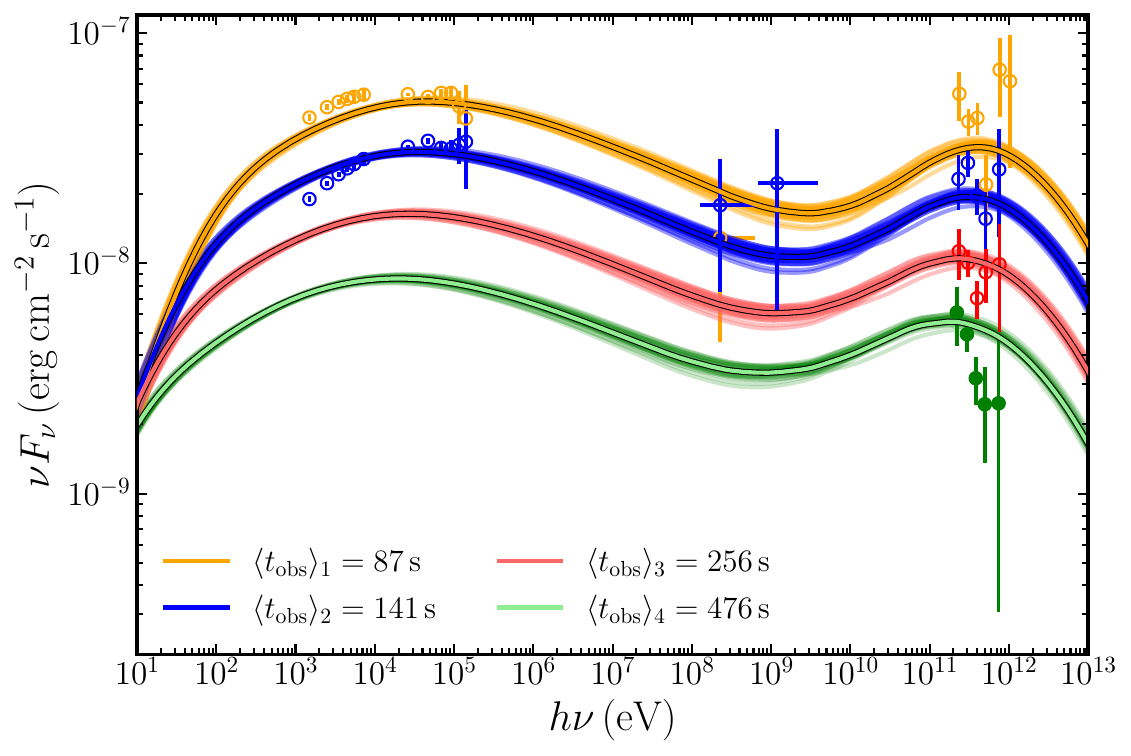} 
\end{minipage}\hfill 
\begin{minipage}[b]{0.50\linewidth}
\centering
s\includegraphics[width=1.\linewidth]{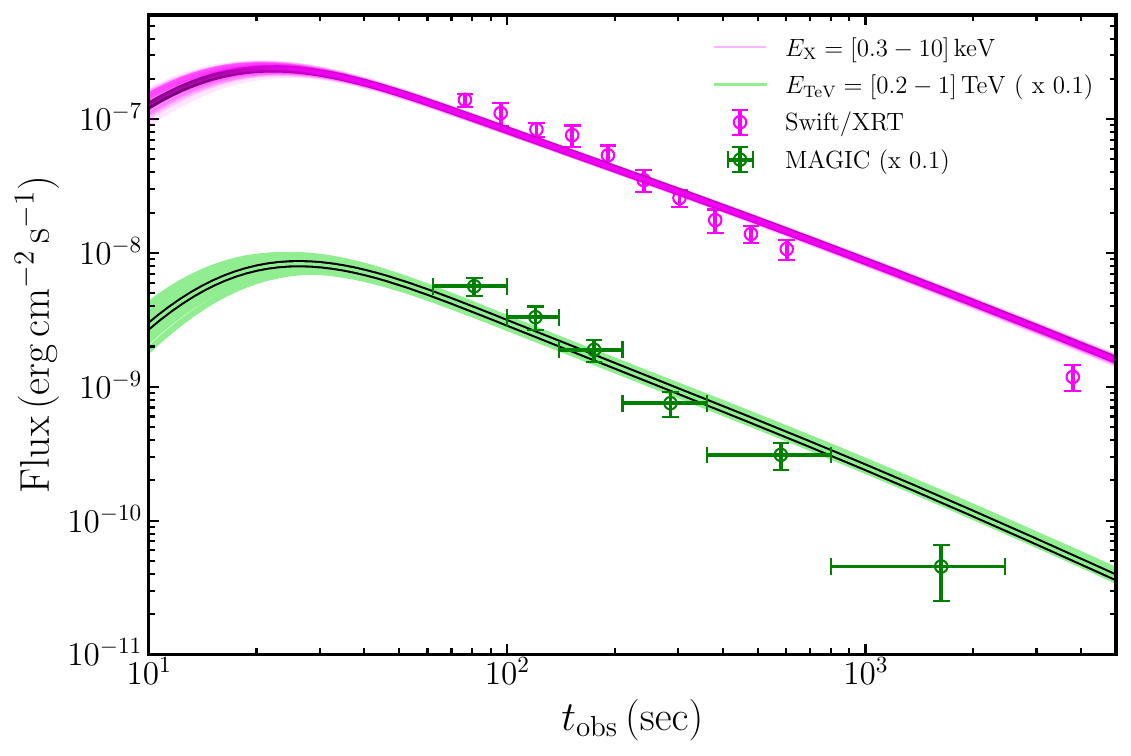} 
\end{minipage} 
\vspace{-16pt} 
\caption{
Spectra and lightcurves corresponding to the best-fit parameters inferred from the MCMC sampling (thick lines) together with solutions randomly sampled from the posterior distribution within the $1\sigma$ credible intervals (thin transparent lines) are shown. Model parameters were obtained fitting the data of three different epochs, $\left\langle t_{\rm obs} \right\rangle$ = 87 s, $\left\langle t_{\rm obs} \right\rangle$ = 141 s and $\left\langle t_{\rm obs} \right\rangle$ = 256 s.
\textbf{(Left)} Spectra at five representative epochs.
\textbf{(Right)} Light curves for integrated flux in the energy band of $E_{\rm X}= [0.3 -10]\,\mathrm{keV}$ and  $E_{\rm TeV} = [0.2$--$1]\,\mathrm{TeV}$.Data were taken from 
https://www.swift.ac.uk
and \citet{MAGIC_GRB190114C}.}
\label{fig_FIT_SPECTRA_LC_3T}
\end{figure*}

\begin{table*}
\centering
\setlength{\tabcolsep}{6pt} 
\renewcommand{\arraystretch}{1.2}
\begin{tabular}{c c c c c c c c c c}
\multicolumn{9}{c}{\textbf{GRB 190114C}} \\
\midrule
  $t_{\rm obs} \, [\rm s]$ & 
  $ E_{k, \rm iso} \, [10^{54} \, \rm erg] $ & 
  $ \Gamma_0 $ & 
  $ n [\rm cm^{-3}] $&
  $ \epsilon_e  \, [10^{-2}] $ & 
  $ \epsilon_B  \, [10^{-4}] $ &  
  $ p $ &  
  $ \xi_e $ &
   \text{Refs.}                %
                   \\
\midrule
 $87$, $141$, $256$  & $7.24_{-2.78}^{+5.06}$ & $575.40_{-38.41}^{+13.40}$ & $0.12_{-0.04}^{+0.08}$ & $2.45_{-0.79}^{+1.01}$ & $8.32_{-2.95}^{+3.98}$ & $2.31_{-0.04}^{+0.04}$ & 1 & \textbf{This work} \\
 $[60-110]$, $[110-180]$& 0.8  & - & 0.5  & 7 & 0.8 & 2.6 & 1 & M19$^{\dag}$\\   
  100 & 0.6  & 300 & 0.3  & 7 & 0.4 & 2.5 & - & W19$^{\dag}$  \\
  80& 1  & 600 & 1.0  & 6 & 9  & 2.3 & 0.3 & A20$^{\dag}$  \\    
  90, 145& 0.3  & - & 2.0  & $\sim$10 & $\sim$27-61 & 2.5 & $\neq 1$ & DP21$^\dag$ \\
  $[66-92]$   & 2.0  & - & $1.0$  & 1 & $5\times10^{-2}$ & 2.3 & 1 & F19$^\dag$  \\
$[110$--$180]$ & 0.8  & 167 & 5  & $1.8^{+2.0}_{-2.0}$ & 42 & $1.6^{+0.4}_{-0.4}$ & - & B25$^{\dag}$ \\
  $[68-110]$ & 0.63  & 500 & $0.2$  & 5 & 50 & 2.8 & - & FT25$^{\dag}$  \\
   $[60-110]$, $[110-180]$ & 10  & 600 & 0.5  & 1 & 0.1 & 2.6 & 1 & N25$^{\dag}$\\   
   $[60$--$110]$, $[110-180]$, & 1.5  & 500 & 1.0 & 10 & 10 & 2.4 & 1 & W26$^{\dag}$\\
   $[180-380]$, $[380-627]$ &   &  &  &  &   &  &  & \\
\midrule
\end{tabular}
\caption{Comparison of our best-fit model parameters with those obtained in different works using $k=0$.  
\\
\textbf{Notes:} M19: \citep{MAGIC_GRB190114C}, W19: \citep{Wang-et-al-19}, A20: \citep{Asano-Murase-Kenji-20}, DP21: \citep{Derishev-Piran-21}, B25: \citep{Barnard+2025}, F19: \citep{ Fraija-et-al-19b, Fraija-et-al-19}, FT25: \citep{Foffano-Tavani-2025}, N25: \citep{Nedora+25}, W26: \citep{Zhao-Feng+26}. We indicate with a $\dag$ those works that either fix the values of some model parameters a priori and/or manually find the best-fit parameters and do not use MCMC.
}

\label{tab:Comparative_GRB190114c}
\end{table*}

We apply our surrogate model and MCMC sampling to fit the broadband afterglow emission of GRB\,190114C \citep{MAGIC_GRB190114C}, a prototype for the class of GRBs showing TeV afterglow emission. We fit its afterglow spectra at three different epochs. Each spectrum corresponds to an average over a finite temporal interval $\Delta t_{\rm obs} = t_{\rm obs, f} - t_{\rm obs, i}$, which is required to achieve sufficient signal-to-noise at high energies, particularly in the TeV regime where photon statistics are limited. Explicitly averaging the model spectra over these intervals would be computationally expensive. Therefore, assuming a temporal evolution of the form $F_\nu \propto t_{\rm obs}^{\alpha}$ with $\alpha \neq 2$, we evaluate the model at an effective time defined by a flux-weighted mean,
\begin{equation}
\langle t_{\rm obs} \rangle = \left( \frac{1 - \alpha}{2 - \alpha } \right)
\frac{t_{\rm obs, f}^{2 - \alpha} - t_{\rm obs, i}^{2 - \alpha}}{t_{\rm obs, f}^{1 - \alpha} - t_{\rm obs, i}^{1 - \alpha}}\,.
\end{equation}
For this calculation, we adopt a temporal decay index value measured in the X-ray band (1--10\,keV), $\alpha_{\rm X} = -1.36 \pm 0.02$ \citep{MAGIC_GRB190114C}. Finally, we impose an additional constraint on the blast-wave dynamics by fixing the lower value of deceleration time to $t_{\rm obs}^{\rm dec}  \simeq (1+z)\frac{r_{\rm dec}}{2c\Gamma_0^2} > 20 $\,s,  corresponding to the epoch when prompt emission terminates and after which both the X-ray and high-energy emission exhibit a smooth power-law decline indicative of the afterglow. Figure\,\ref{fig_FIT_SPECTRA_LC_3T} shows our best-fit (corresponding to the median values of the parameter posterior distributions, as shown in Fig.\,\ref{fig:corner_GRB190114C}) model spectrum and light curve, along with the same produced using 30 random samples of the parameter posterior distributions (see Fig.\,\ref{fig:corner_GRB190114C}) that illustrate the spread in our fit due to inferred parameter uncertainties. We obtained this solution by performing a joint fit of the MAGIC observations at $\left\langle t_{\rm obs} \right\rangle = 87 \, \rm s$, $\left\langle t_{\rm obs} \right\rangle =141 \, \rm s$ and $\left\langle t_{\rm obs} \right\rangle =256 \, \rm s$. The best-fit parameters obtained in this work for GRB 190114C are summarized in Table~\ref{tab:Comparative_GRB190114c}, together with values reported in previous studies that also only explored solutions for an ISM environment. 

Overall, our results are broadly consistent with earlier SSC interpretations of the GRB 190114C afterglow emission, particularly in the requirement of a highly relativistic outflow in the range of $\Gamma_0 \sim 500$--$600$ and electron spectral indices in the range $p \sim 2.3$--$2.6$. These values are compatible with the standard picture of relativistic external shocks powering the broadband afterglow emission.
Among the previous studies, the results of \cite{Barnard+2025} deviate most noticeably from the values reported by other works, featuring a particularly hard electron energy distribution ($p=1.6$) and a substantially lower initial Lorentz factor ($\Gamma_0 = 167$).

We find the isotropic-equivalent kinetic energy, $E_{k,\rm iso} = 7.24^{+5.06}_{-2.78}\times10^{54}$\,erg, to be higher than most previous works by a factor of $\sim4-23$, but it is more consistent with the results of \citet{Nedora+24} and \citet{Aguilar-Ruiz+26} who used a code similar to our code.
The inferred external density, $n_0 = 0.12^{+0.08}_{-0.04} \, \rm cm^{-3}$, favors a relatively tenuous circumburst environment for a collapsar GRB, and it broadly agrees with several earlier works that assume an ISM environment. 

The electron energy fraction is constrained to $\epsilon_e = 2.45^{+1.01}_{-0.79}\times10^{-2}$. This value is lower than several of the estimates from previous studies listed in Table \ref{tab:Comparative_GRB190114c}.
A lower $\epsilon_e$ implies that a smaller fraction of the dissipated shock energy is transferred to non-thermal electrons. Consequently, the observed luminosity must be compensated by a larger kinetic energy reservoir, consistent with the high value of $E_{k,\rm iso}$ obtained in the present analysis. The inferred magnetic energy fraction, $\epsilon_B = 8.32^{+3.98}_{-2.95}\times10^{-4}$, is moderate and lies within the broad range of values reported in previous works, which spans approximately three orders of magnitude, from $5 \times 10^{-6}$ to $6 \times 10^{-3}$. Combined with $\epsilon_e \simeq 2.45\times10^{-2}$, our solution implies $\epsilon_e/\epsilon_B \approx 30$. Such a configuration, in which the post-shock energy budget is dominated by relativistic electrons rather than magnetic fields, naturally favors efficient SSC emission, with a Compton Y-parameter of order unity to several, consistent with the bright very-high-energy component detected by MAGIC.

Figure \ref{fig_FIT_SPECTRA_LC_3T} shows that although the best-fit parameters successfully reproduce the broadband spectra at early epochs, they partially fail to capture the late-time behavior of the afterglow. This discrepancy becomes evident in the light curves, where the model increasingly deviates from the observations beyond $\sim10^{3}$ s. Such behavior may indicate that the assumption of a homogeneous external ISM ($k=0$) is overly simplistic and that a more complex external-density structure is required. According to the standard framework of relativistic external shocks propagating into a constant-density medium, the temporal decay index $\alpha = d\ln F_\nu/d\ln t_{\rm obs}$ and the electron energy distribution power-law index $p$ are related through the synchrotron closure relations \citep{Granot-Sari-02}. Assuming that the X-ray band lies above the cooling frequency ($\nu>\nu_c$), the monochromatic flux evolves as $F_\nu \propto t_{\rm obs}^{-(3p-2)/4}$. Adopting the observed X-ray decay slope $\alpha_{\rm X}= - 1.36$ yields an implied electron index of $p \simeq 2.48\pm0.03$, which is slightly larger than our inferred value of $p=2.31^{+0.04}_{-0.04}$. This tension may indicate limitations of the simple $k=0$ scenario when broadband spectral and temporal constraints are considered simultaneously. Indeed, the discrepancy can be alleviated by adopting a more general stratified external medium, as discussed by \citet{Aguilar-Ruiz+26}, who found improved agreement between the spectral and temporal evolution of the afterglow. In order to constrain the external-density profile simultaneously with the other physical parameters, we will extend the model grid to include $k$ as a free parameter in future work (Aguilar-Ruiz et al., in prep.).

Most previous studies assuming a constant-density environment did not perform simultaneous multi-epoch spectral fits. The exceptions are \citet{Nedora+25} and \citet{Zhao-Feng+26}, who modeled two and four epochs, respectively, although neither study presented a detailed comparison with the late-time lightcurve evolution. 
Notably, \cite{Barnard+2025} compiled data from ten observational epochs, however, their MCMC analysis was performed using only a single epoch at 110-180 s.
In contrast, e.g., \citet{Foffano-Tavani-2025} find a good match between the model and the observations in the X-ray and TeV gamma-ray bands, however their resulting broadband spectrum in the interval time of $68\leq t_{\rm obs}\,[{\rm s}]\leq 110$ shows disagreement in the X-ray component. Similar results are found in \cite{Asano-Murase-Kenji-20} whose model exhibit good agreement with the spectrum at $t_{\rm obs}=80$\,s, while the X-rays and TeV gamma-ray lightcurves indicate deviations from data at later times. On the other hand, \citet{Derishev-Piran-21} fitted two epochs independently, requiring different sets of physical parameters at each epoch, particularly for the microphysical fractions $\epsilon_e$ and $\epsilon_B$.

Furthermore, previous studies report a wide range of inferred physical parameters for GRB 190114C. Despite these differences, most published solutions imply deceleration times in the range of approximately $t^{\rm dec}_{\rm obs} \sim 5-40 \, \rm s$. This suggests that different combinations of $E_{k, \rm iso}$, $n_0$, and $\Gamma_0$ can lead to a broadly similar dynamical evolution of the external shock. The posterior distributions from our analysis show strong correlations among $E_{k, \rm iso}$, $n_0$, $\epsilon_e$, and $\epsilon_B$. Together with the wide range of parameter values reported in previous studies, these correlations highlight the need for caution when comparing individual parameter estimates between different analyses.

\section{Summary \& Discussion}\label{sec:Discussion}

We developed an ANN surrogate model for emulating broadband SSC afterglow emission from a blast wave propagating inside a constant density ISM external medium ($k=0$). To train the ANN we used our numerical code that solves the integro-differential kinetic equations accounting for all relevant interactions between electrons and photons. For the on-grid validation sample, the surrogate reproduces the original spectra to within a few per cent across most of the modeled energy range while reducing the computational cost by several orders of magnitude. The off-grid tests reveal a localized interpolation failure at $2.01<p\lesssim 2.18$, demonstrating that on-grid validation alone does not fully characterize interpolation accuracy throughout the continuous parameter domain. Both applications presented here lie outside this affected region. 

The systematic increase in the REs below the IR bands ($h\nu_* \lesssim 0.1 \, \rm eV$) and above the ultra-high-energy (UHE) gamma-ray regime ($h\nu_* \gtrsim 100 \, \rm TeV$) is a result of the energy grid in the comoving frame being narrower than that in the observer frame. The manual extrapolations of the spectrum applied at both ends in the observer-frame introduce small inaccuracies that lead to significant REs.
Thus, $0.1 \,\rm eV$--$100 \,\rm TeV$ can be considered the most reliable energy range of our model. 
When examining the 97.5th percentile RE, we found that an additional limitation arises near the transition between the synchrotron and SSC components, approximately in the $10 \,\rm GeV$--$1 \,\rm TeV$ range, where the REs can increase to $\sim 12\%$ at early times, particularly between $t_{\rm obs*} =10-100 \, \rm s$. This limitation is particularly relevant for TeV observations, which are typically restricted to $t_{\rm obs}<10^4$ s as the emission becomes too faint to be detected at later times. This sets a practical accuracy limit for the surrogate model in the energy range most relevant to very-early-time TeV observations. Reducing these errors would require a finer grid with additional nodal points, which would substantially increase the computational cost.
Nevertheless, these larger REs at early times do not undermine the overall applicability of the model, as the resulting precision remains within the observational uncertainties typically associated with the VHE gamma-ray band measured by current facilities.

The surrogate model enables rapid forward-model evaluations, reducing the computational evaluation time by several orders of magnitude compared to full numerical simulations. This acceleration makes conventional likelihood-based Bayesian inference using MCMC sampling computationally feasible for fitting light curves and broadband spectra. To account for the finite precision of the surrogate model, we include an additional term in the log-likelihood function,  $\sigma_{\rm NN}$, based on the 97.5th percentile of the relative-error distribution in the on-grid validation sample for each time decade.

We applied our surrogate model together with MCMC sampling to the prototypical TeV GRB 190114C. Our results indicate an overall good match of the broadband spectral behavior, however, the modeled lightcurves exhibit small discrepancies at later times, particularly for $t_{\rm obs} \gtrsim 10^3$ s (see Fig. \ref{fig_FIT_SPECTRA_LC_3T}). The inferred parameters favor a high isotropic-equivalent kinetic energy and a relatively tenuous circumburst environment, together with $\epsilon_e / \epsilon_B \sim 30 $, yielding conditions that naturally support efficient SSC emission and are consistent with the bright very-high-energy component detected by MAGIC. 

Most previous modeling efforts for this burst often relied on single-epoch snapshots or manual parameter tuning. In contrast, we self-consistently perform a simultaneous multi-epoch spectral fit. The deviation of our model light curve, in particular at $\sim\rm{TeV}$ energies, from observations at late times suggests that the assumption of an ISM external medium is not correct, and that the external medium should have a radial profile. The X-ray emission arises from the synchrotron power-law segment with $\nu_m<\nu_X<\nu_c$ and the flux density in this segment decays with a slope $\alpha=[k(3p-5)-12(p-1)]/4(4-k)$. Furthermore, the cooling break frequency that enters the \textit{Swift}/XRT range declines with time with a temporal slope $d\ln\nu_c/d\ln t_{\rm obs}=(3k-4)/(8-2k)$. Therefore, the X-ray model light curve will not produce the correct decay trend if the true value of $k\neq0$. The TeV emission arises near the peak of the SSC spectrum, which in the simpler Thomson regime is produced by the IC scattered synchrotron cooling-break photons. Therefore, the SSC peak must also depend on the value of $k$. A more complete treatment by \citet{Aguilar-Ruiz+26} that allowed for a varying $k$ showed that the observations indeed prefer a value of $k=1.67$, which is closer to a wind external medium as expected in a collapsar GRB.

Looking forward, the clear necessity for wider grid parameter space, extending to stratified environments ($k \neq 0$) to simultaneously achieve early-time spectral consistency and late-time light curve represents the immediate evolution of this framework. Furthermore, although the results obtained in this work are restricted to the emulation of spherical outflows, they provide a promising foundation for extending our calculations to jets with angular structure. Such an extension would benefit from a computationally efficient approach, such as the semi-analytic formalism proposed by \citet{Aguilar-Ruiz+26}, which can significantly reduce the computational cost of generating the training dataset compared to fully numerical kinetic calculations. Combining these approaches would allow us to explore more complex jet structures while retaining the accuracy and computational efficiency required for parameter inference over large parameter spaces. This will be a topic of a future work (Aguilar-Ruiz et al. in preparation). Such extensions will enable comprehensive population studies of TeV GRBs and facilitate rapid interpretation of future observations from facilities such as the Cherenkov Telescope Array Observatory and other next-generation high-energy observatories.

\section*{Acknowledgements}

EAR acknowledges support from the SECIHTI postdoctoral fellowship.

\section*{Data Availability}
The model grid used to train the neural-network surrogate presented in this work is available on Zenodo at \url{https://doi.org/10.5281/zenodo.23006149}. The dataset contains the 100\,800 model spectra described in Section \ref{subsec:gridgen} and is provided in ASDF format.



\bibliographystyle{mnras}
\bibliography{refs} 



\newpage
\clearpage

\appendix

\section{Fit Priors and Posterior Distributions}

\begin{table}
\centering
\caption{Prior ranges adopted in the MCMC sampling. All parameters are sampled with flat priors within the listed intervals. The chosen limits are given by the limits values of the grid model. An additional constraint was imposed $t_{\rm obs}^{\rm dec} >20 \, \rm s$.
}
\begin{tabular}{lcc}
\hline

\hline
Parameter & Distribution  & Range\\
\hline
$\log E_{k, \rm iso}$ & uniform & [52, 55]\\
$\log\Gamma_0$ & uniform& [$\log(100), \log(600)$] \\
$\log n$ & uniform & [-1, 1]\\
$\log\epsilon_e$ & uniform & [-3, -1]\\
$\log\epsilon_B$ & uniform & [-5, -2]\\
$p$ & uniform & [2.01, 2.6] \\
$t_{\rm obs}^{\rm dec}$ & None & $>20$\,s \\
\hline
\end{tabular}
\label{tab:relative_errors}
\end{table}

\begin{figure*}
    \centering
    \includegraphics[width=1.\linewidth]{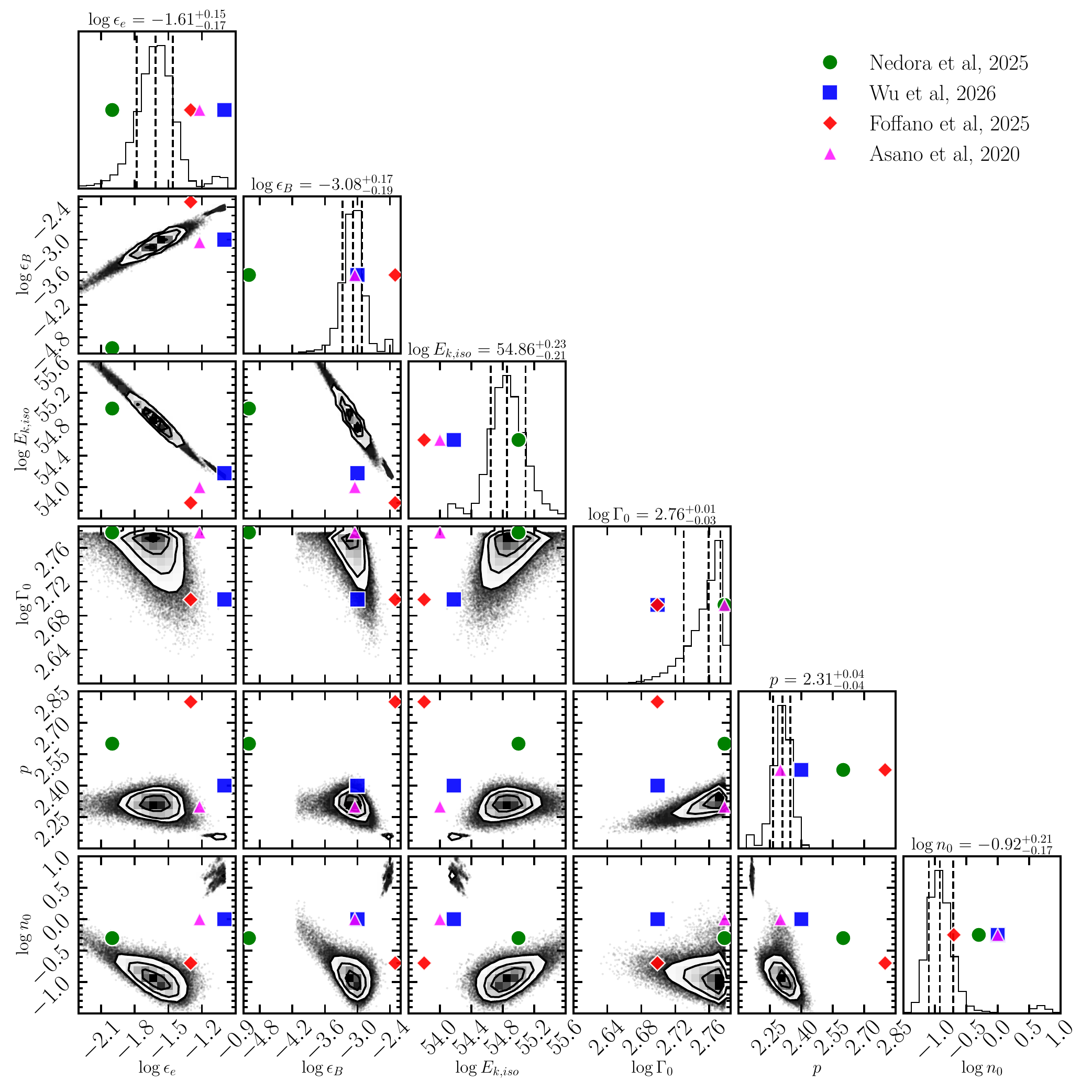}
    \caption{Posterior probability distributions of the model parameters obtained through an MCMC sampling of the broadband afterglow data of GRB 190114C. The diagonal panels display the marginalized distributions for each parameter, while the vertical dashed lines mark the median (50th percentile) and the 16th and 84th percentiles 
    For comparison we shows the parameter values obtained by other works which uses a kinetic approach and $k=0$ for a spherical flow, see Table \ref{tab:Comparative_GRB190114c}. 
    }
    \label{fig:corner_GRB190114C}
\end{figure*}

\clearpage

\section{MCMC fit to synthetic data}

\begin{table}
\centering
\caption{Relative errors between true values and mean inferred values for each parameter for MCMC fits of synthetic data.}
\begin{tabular}{lc}
\hline
Parameter & Relative Error (\%) \\
\hline
$\epsilon_e$ & 2.50 \\
$\epsilon_B$ & 0.46 \\
$E_{k, \rm iso}$ & 9.90 \\
$\Gamma_0$ & 1.60 \\
$p$ & 0.08 \\
$n$ & 0.93 \\
\hline
\end{tabular}
\label{tab:relative_errors}
\end{table}

\begin{figure*}
    \centering
    \begin{minipage}[b]{1.\linewidth}
        \centering
        \includegraphics[width=0.8\linewidth]{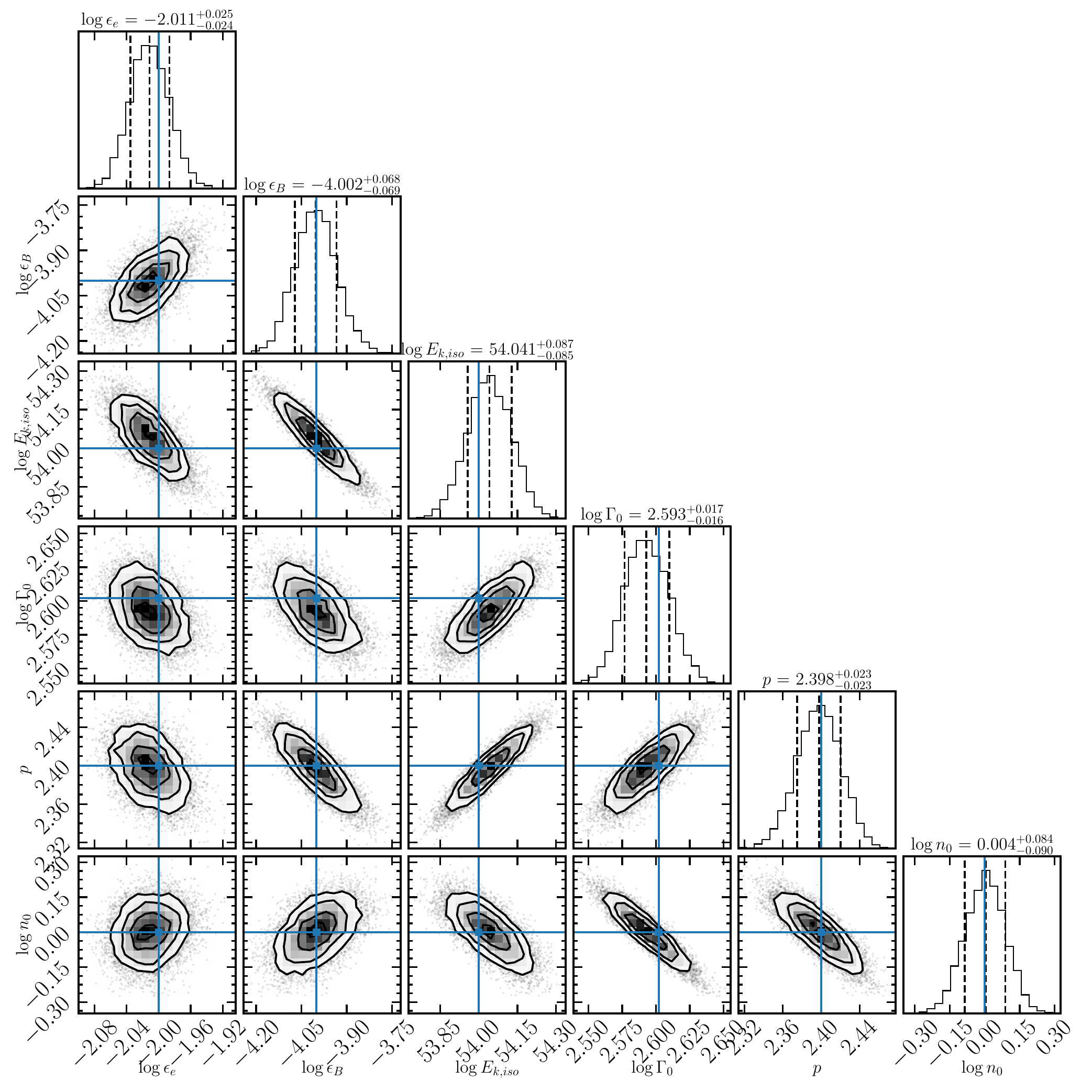}
    \end{minipage}\hfill 
    \begin{minipage}[b]{1.\linewidth}
        \centering
        \includegraphics[width=0.6\linewidth]{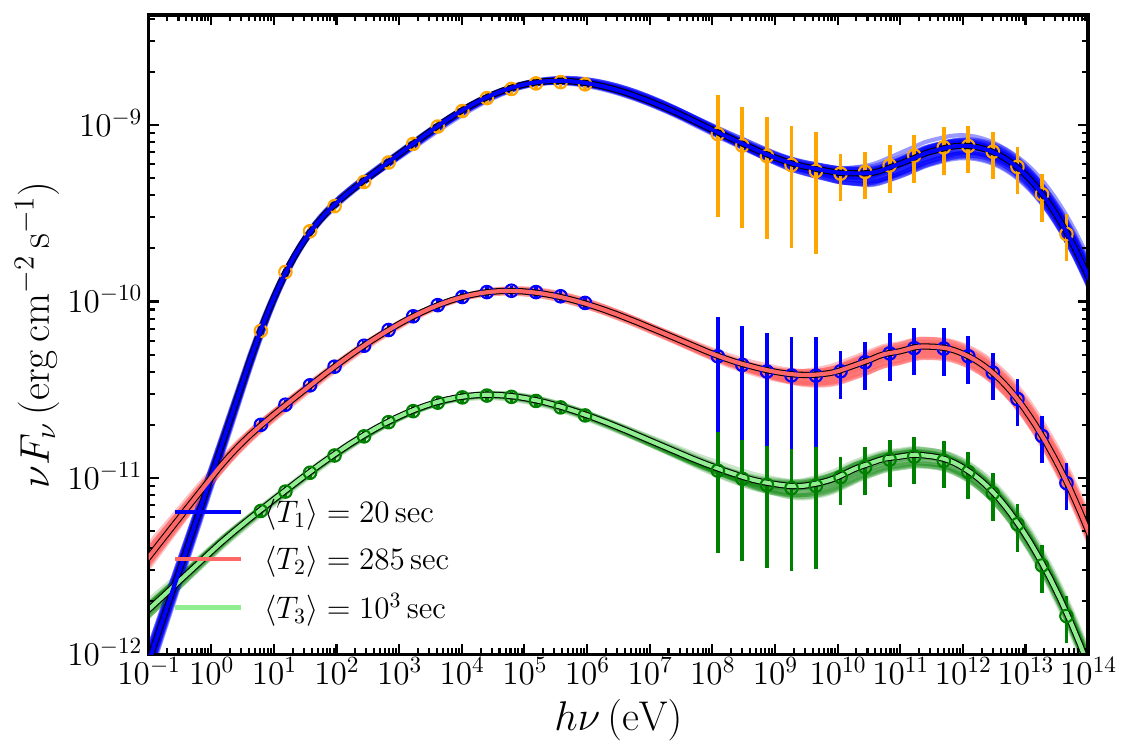}
    \end{minipage} 
   \vspace{-16pt} 
    \caption{ The top panel shows the marginal posterior distributions obtained from our MCMC analysis of the synthetic data. The true parameter values are marked with blue crosses, while the vertical lines indicate the median and the $1\sigma$ credible intervals.  
The bottom panel shows the synthetic data generated using the kinetic formalism described in this work and the following model parameters: 
$
\epsilon_e = 10^{-2}, \; 
\epsilon_B = 10^{-4}, \;
E_{k, \rm iso} = 10^{54} \, {\rm erg,} \;
\Gamma_0 = 400, \;
p = 2.4, \;
n = 1 \, {\rm cm^{-3}}, \;
z = 0, \;
d_L = 10^{28} \, {\rm cm},
$ 
along with the set of solutions within the $1\sigma$ credible interval.
}
    \label{fig_CORNER_synthetic}
\end{figure*}

\label{lastpage}
\end{document}